\documentclass[12pt]{spieman}  
\pdfoutput=1
\usepackage{amsmath,amsfonts,amssymb}
\usepackage{graphicx}
\usepackage{setspace}
\usepackage{tocloft}
\title{Predictive model of persistence in H2RG detectors}

\author[a*]{Simon Tulloch}
\author[a]{Elizabeth George}
\author[a]{ESO Detector Systems Group}
\affil[a]{ESO, Karl Schwarzschild Stra{\ss}e 2, Garching, Germany, 85748}

\cftpagenumbersoff{figure}
\cftpagenumbersoff{table}
\begin{document}
\maketitle

\begin{abstract}
Infrared hybridized detectors are widely used in astronomy, and their performance can be degraded by image persistence.  This results in remnant images that can persist in the detector for many hours, contaminating any subsequent low-background observations. A different but related problem is reciprocity failure whereby the detector is less sensitive to low flux observations. It is demonstrated that both of these problems can be explained by trapping and detrapping currents that move charge back and forward across the depletion region boundary of the photodiodes within each pixel. These traps  have been characterized in one 2.5 $\mu$m and two 5.3 $\mu$m cutoff wavelength Teledyne H2RG detectors. We have developed a behaviour model of these traps using a 5-pole Infinite Impulse Response digital filter. This model allows the trapped charge in a detector to be constantly calculated for arbitrary exposure histories, providing a near real-time correction for image persistence.
\end{abstract}


{\noindent \footnotesize\textbf{*}Simon Tulloch,  \linkable{smt@qucam.com} }

\begin{spacing}{1}   

\section{Introduction}
\label{sect:intro}  
In the last 20 years, large hybridized mercury cadmium telluride (MCT) infrared arrays have become standard detectors used in astronomy. As the noise performance and uniformity of these arrays has improved, subtle effects that can bias astronomical results have emerged as important components in the error budgets of the detector systems in astronomical instruments. One of the most problematic of these effects is image persistence, the remnant charge that remains in the detector after an observation, which appears as an ``after-glow'' in subsequent images.

In 2008, Smith et al. \cite{SMITH1,SMITH2} hypothesized that image persistence is due to traps in the depletion region of the diode. Since that time, the model has been largely adopted by the community, and much work has been done on characterizing traps in infrared detectors. Anderson et al.\cite{Anderson} used electrical stimulation of the detector to map out the physical location of traps within each pixel, and illumination tests done by Regan et al.\cite{Regan} provided data supporting the theory that these traps are also at least partially responsible for reciprocity failure (also known as count rate nonlinearity), though Biesiadzinski et al. \cite{Biesiadzinski} show that the traps cannot entirely explain reciprocity failure.

Many authors have worked to characterize the appearance and decay of persistence as a function of flux and fluence to develop phenomenological models to correct persistence in astronomical data. Mosby et al.\cite{Mosby} and Leisenring et al. \cite{leisenring} characterized time constants of persistence decay in the SALT RSS-NIR and JWST NIRCam detectors and found that 2- and 3-exponential decay models respectively described the persistence decay well, potentially pointing to 2 or 3 distinct trap populations.  Other phenomenological models exist indicating that a power-law decay model fits the data well \cite{baril}\(^,\)~\cite{long} indicating that a large range of trapping time constants are responsible for the observed persistence.

Modelled behavior of image persistence has been used to plan observations and place operational restrictions on what astronomical sources may be observed,\cite{Mosby}\(^,\)~\cite{Crouzet} and the ultimate goal is to model persistence accurately enough so that it may be corrected for in astronomical images. To our knowledge, the only instrument currently implementing a standard correction for image persistence is the WFC3cam on Hubble, which uses the model developed in Long et al.\cite{long}. The correction model is based on the power-law decay model with a limited range of time validity, and a user who suspects image persistence is contaminating their data must request that a persistence map be specifically created for their data, requiring human intervention for each dataset that must be corrected. WFC3cam users also have access to a MAST tool that warns of any preceding images that may have left persistence signatures.

We have built on this previous work to develop a standardized trap characterization method and automatic near real-time persistence correction algorithm valid for all times. In Sec.~\ref{sect:theorytraps} , we describe the theory of traps in MCT. Our characterization method probes the behavior and number of traps in the MCT on a per-pixel basis, described in Sec.~\ref{sect:character}. In Sec.~\ref{sect:results}, this method was applied to two 5.3\(\mu\)m cutoff and one 2.5\(\mu\)m cutoff H2RG detectors.  The characterization information was then used to construct a fast persistence correction algorithm based on digital filters that keeps a record of the trapped charge in each pixel at all times, described in Sec.~\ref{sect:predict}. This allows a model persistence/reciprocity failure map to be automatically generated at the time of every exposure and appended to the data as a fits extension.  In Sec.~\ref{sect:accuracy} the model is validated using astronomically realistic exposures. We conclude in Sec.~\ref{sect:concs} and propose future avenues of research in Sec.~\ref{sect:future}.

\section{Theory of Traps}
\label{sect:theorytraps}
\begin{figure}[htbp]
\begin{center}
\begin{tabular}{c}
\includegraphics[height=5cm]{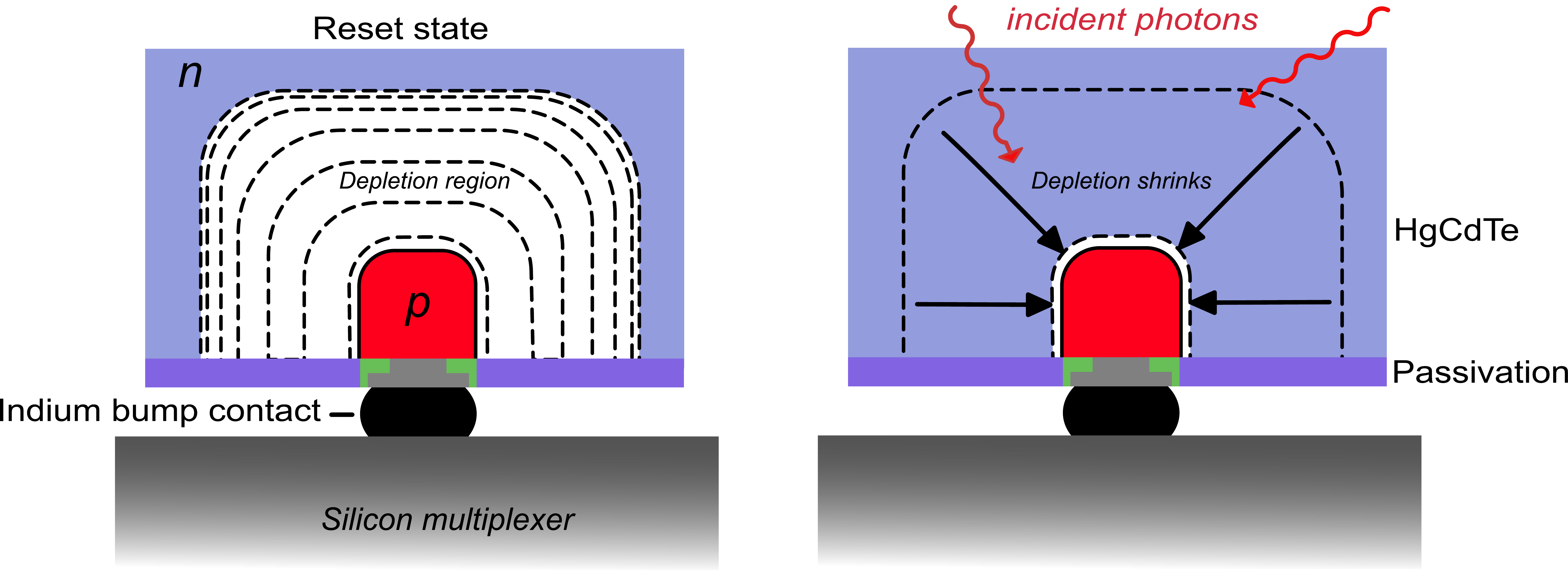}
\end{tabular}
\end{center}
\caption
{ \label{fig:pixel}
Hybrid pixel schematic. The H2RG photo-diode array is implemented in a wafer of MCT material bump bonded to a Silicon multiplexer. A pixel is shown in its reset state on the left and fully exposed on the right.}
\end{figure}

To understand persistence one must first consider the geometry of a hybrid pixel. This is shown in Fig.~\ref{fig:pixel}. At the start of an observation the pixel is reset, expanding to  a maximum the volume of its depletion region (DR) . After reset, incident light generates electron-hole pairs that progressively discharge the pixel and cause the DR to shrink, reaching  minimum volume at full-well. Some of these charge carriers fall into defects in the semiconductor lattice, known as traps, where they are temporarily captured. They are later released with a very wide range of characteristic time constants. Smith\cite{SMITH1} has hypothesized that traps within the DR are responsible for persistence. The current study supports this idea.

\begin{figure}[htbp]
\begin{center}
\begin{tabular}{c}
\includegraphics[height=4cm]{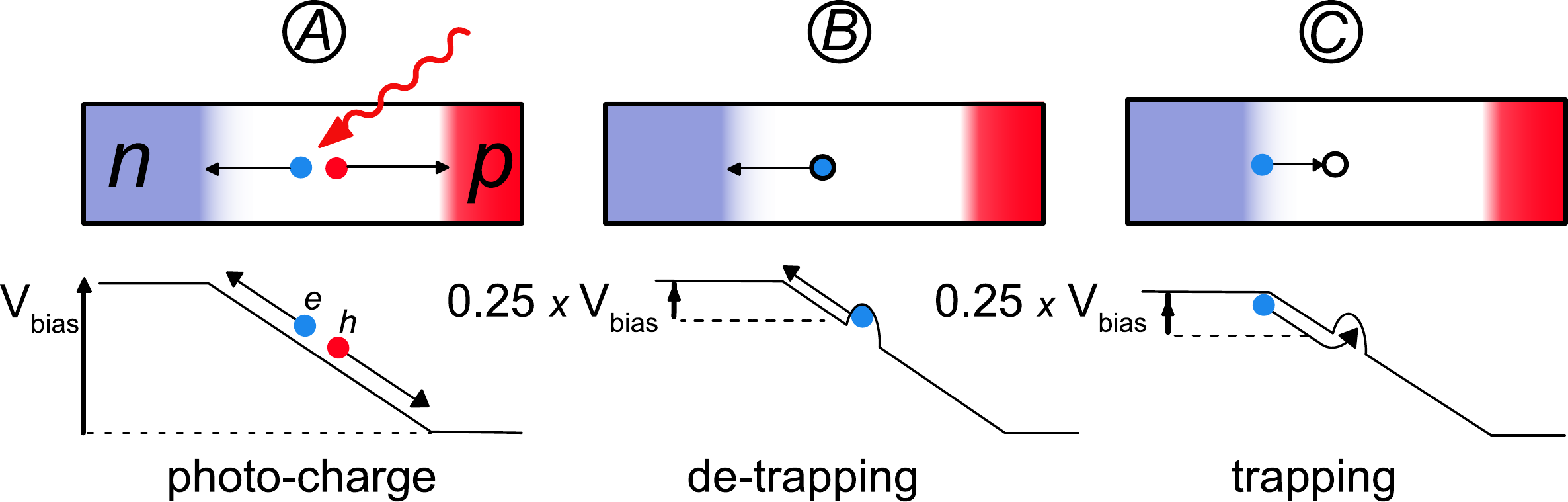}
\end{tabular}
\end{center}
\caption
{ \label{fig:depletionModel}
Charge movement to and from the DR. Trapping and detrapping charge carriers have fractional effective charge. This example shows the behaviour of a trap that is 1/4 of the way across the depletion region. In reality the trap positions are widely distributed. }
\end{figure}

\begin{figure}[htbp]
\begin{center}
\begin{tabular}{c}
\includegraphics[height=12cm]{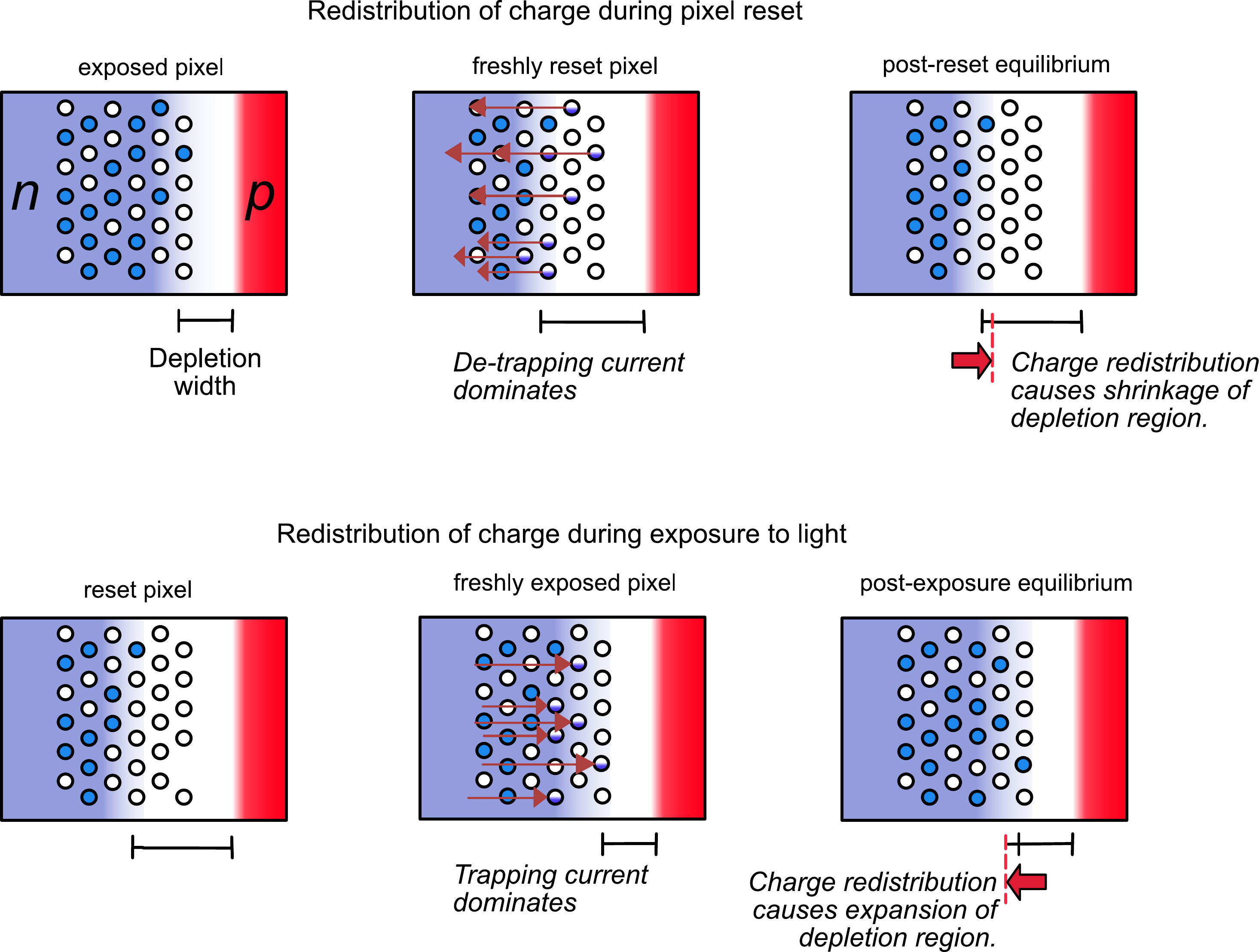}
\end{tabular}
\end{center}
\caption
{ \label{fig:chargemovement}
Movement of charge across depletion region edge in case of a pixel reset (top image row) and during image integration (bottom image row). Empty circles represent trap sites, full blue circles represent traps containing charge. }
\end{figure}

Movement of charge in and out of traps does not by itself affect the measured value of a pixel. It is only if this charge moves through an electric field that we can observe a change in pixel voltage. The \textit{n} and the \textit{p} regions of the photo-diodes are conductive and so have very low electric fields. The DR on the other hand is an insulator and contains a strong electric field. Panel A of Fig.~\ref{fig:depletionModel} shows how a photo-electron/hole pair generated in the DR is swept apart by this field. Energetically this is equivalent to a single electron moving through the full width of the junction i.e. through the full bias voltage on the diode. Panel B shows an electron being released from a DR trap and moving to the \textit{n} region where it adds signal to any photo-charge that may already be present. Since in this example the detrapped charge only moves through 1/4 of the electric field, it will generate only 1/4 of the signal of a photo-electron. Panel C shows the opposite process whereby an electron moves from the \textit{n} region to the DR trap. This trapping process will, in this particular example, produce a reduction in pixel voltage equal to 1/4 of a photo-electron although in reality the traps will be distributed widely throughout the semiconductor material. Fig.~\ref{fig:chargemovement} shows the hypothesized trap behavior as the exposure level of the pixel varies.

 Starting at top left, a pixel is shown in its exposed state. In this equilibrium state there are an equal number of traps filled per unit time as are detrapped. Only a small fraction of the traps within the doped \textit{n} region are filled, as are an even  smaller fraction of the traps lying within the DR. Moving to the right the pixel is then reset causing a shrinkage of the \textit{n} region. A number of filled traps then find themselves within the DR from where they slowly emit their charge with characteristic time constants. Previously filled DR traps are unlikely to be refilled once they emit their charge as their increased distance from the depletion edge means such transitions have a low probability. Any emitted charge then moves to the \textit{n} region where it causes an increase in signal level that mimics the effect of newly received photo-charge i.e. causes persistence. A new equilibrium is eventually found where trapping and detrapping current across the depletion edge are once again equal and opposite. Note that a reset will also shrink very slightly the \textit{p} region and presumably cause a redistribution of trapped holes. Due to the higher p-doping concentration, this shrinkage will be very much smaller than that experienced by the \textit{n} region and the effect of these holes has not been considered in this study. Moving to the bottom left of Fig.~\ref{fig:chargemovement} we find a pixel in its reset equilibrium state. The DR is expanded to its full width and a fraction of the traps within the \textit{n} region are filled. Moving to the right we then see the effect of illumination. The DR shrinks in direct response to the generated photo-charge. The distribution of filled traps then  seeks a new equilibrium. This produces a net transfer of charge to the DR with a characteristic time constant causing a ``rebound'' in its width and a partial reset i.e. a loss of signal. Another way of thinking about the rebound effect, both following a step in exposure level and also a reset event, is that the pixel takes some time to respond to the external stimulus. In both cases the slow redistribution of charge in the depletion region acts to oppose this stimulus.
 Characterizing persistence in a detector requires the number of traps and their trapping and detrapping time constants to be measured.

 \begin{figure}[htbp]
\begin{center}
\begin{tabular}{c}
\includegraphics[height=5.5cm]{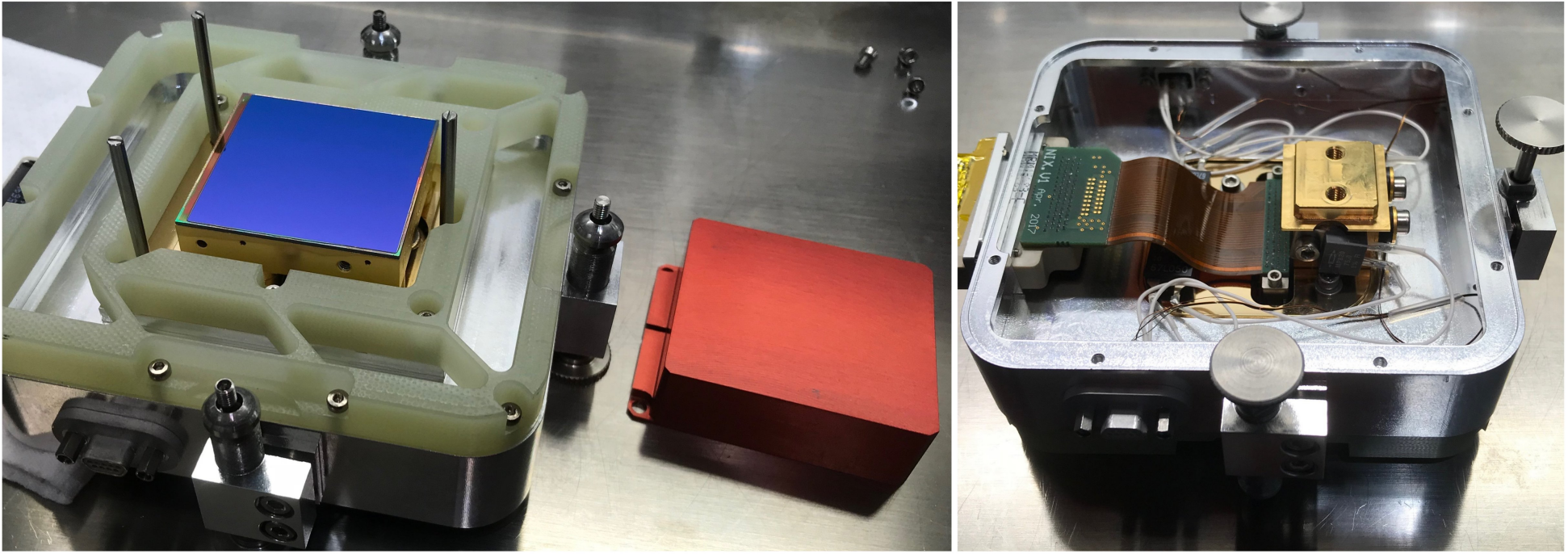}
\end{tabular}
\end{center}
\caption
{ \label{fig:Mount}
Detector mount used for characterisation of the ERIS H2RG sensors. }
\end{figure}

\section{Trap Characterization}
\label{sect:character}
 Three H2RG devices were characterized at ESO in a low-background cryostat. The first was a 5.3\(\mu\)m sensor delivered to ESO before 2013, used in the CRIRES+ project \cite{CRIRES+}. The second and third were 2.5\(\mu\)m and 5.3\(\mu\)m devices delivered in 2018 and destined for the ERIS project \cite{ERIS}. The test cryostat was capable of cooling to 37K. The long-wavelength detectors were operated at 40K to reduce dark current and improve cosmetics. The short-wavelength detector was operated at 80K since its final instrument will be LN$_{2}$ cooled. The test cryostat contained a black body and filter wheel for illuminating the detectors at J,H,K,L and M bands as well as a 900nm LED capable of generating low-uniformity flat fields. The LED was driven from a spare clock in the detector controller and could be pulsed synchronously to the exposure and readout processes.

 Persistence measurements consisted essentially of accurate photometry of a small previously illuminated window. A number of noise sources had to be dealt with. Bias drift was present in the DC-coupled video processor chain. This was removed by doing subtraction of the reference pixels in both the x and y axes. Aligning the 64 x 64 measurement window to a single H2RG amplifier also helped reduce the effects of bias drift. Dark current and instrumental background was also a big problem particularly in the ERIS 5.3\(\mu\)m devices. These two noise sources were mitigated for the two ERIS devices by interspersing persistence measurement runs with dark reference runs where the LED was switched off whilst all other conditions remained the same. The CRIRES+ device had very low background and also low intrinsic dark current. It had the additional advantage of having large areas shadowed from the LED which meant that a dark current reference window could be read simultaneously to the persistence measurement window. This device therefore gave the best quality data. The cosmetics of the H2RG devices used were  poor although within specification. Large numbers of hot and dead pixels were present in all parts of the images and many fell unavoidably into our measurement windows. These pixels were rejected by measuring the median window pixel value rather than the mean. Towards the end of the study a new technique of artificially generating dark reference windows at arbitrary positions on the devices using selective and repetitive line resets during exposure was developed.

\label{sect:characterization}  




\subsection{Probing Trap Time Constants}
\label{sect:probetau}
\begin{figure}[htbp]
\begin{center}
\begin{tabular}{c}
\includegraphics[height=6cm]{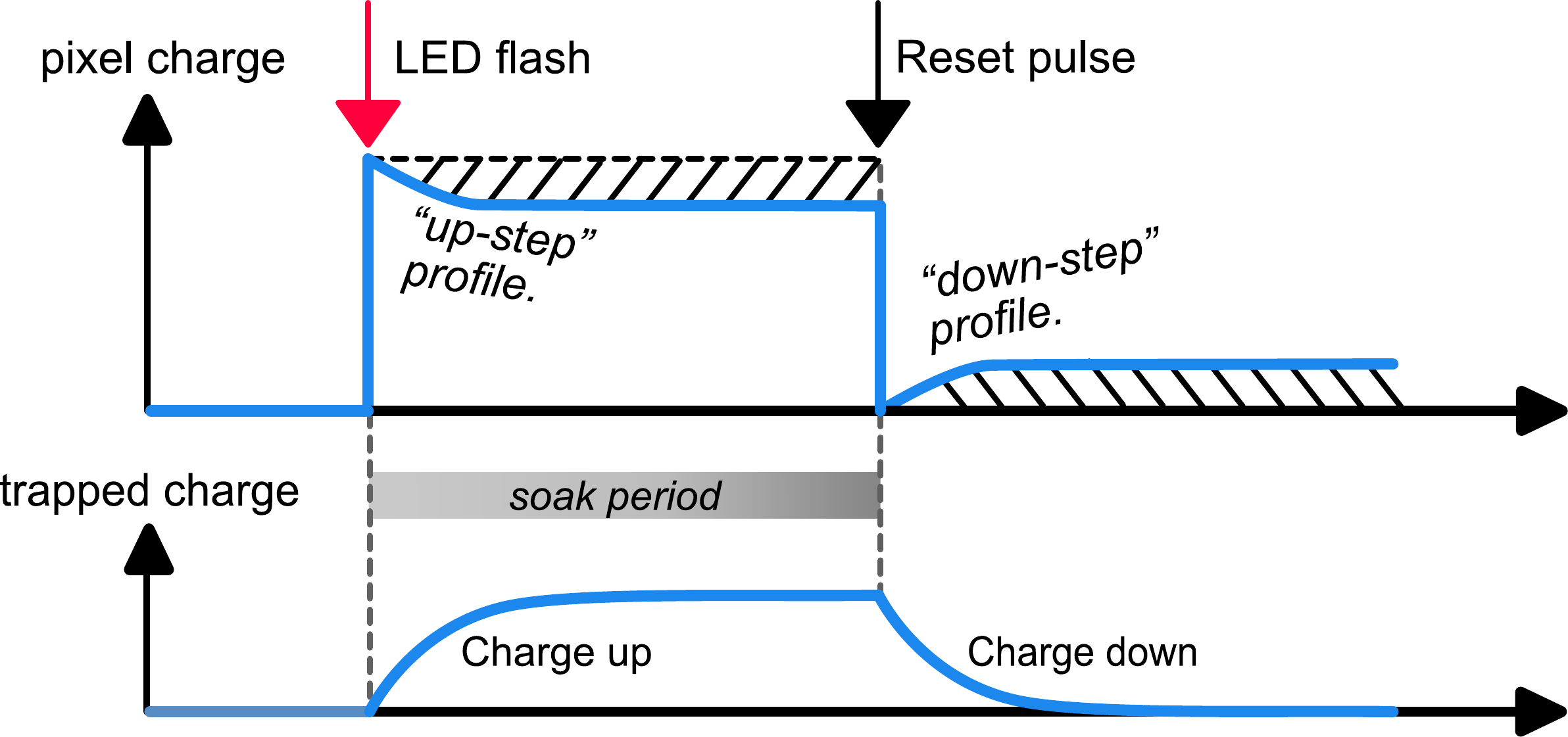}
\end{tabular}
\end{center}
\caption
{ \label{fig:LEDmethod}
The LED method used to probe the trapping and de-trapping time constants. Key parts of the trapped charge and signal level profiles are shown labeled.}
\end{figure}

The behaviour of traps clearly had a lot in common with that of a low pass electronic filter whose simplest form is an RC network.
In the case of a real RC network, the time constants can be experimentally determined by applying electrical delta functions, sine waves, or step functions and observing the output response. Only the last of these is really a practical option for probing time constants in an H2RG pixel. A step function can be generated by first applying an LED pulse to expose the pixel to a determined level, seeing how this level then decreases due to trapping over a certain time, followed by a reset of the pixel and a measurement of how this trapped charge then slowly reappears in the subsequent image as persistence. The LED was bright enough to give a full-well exposure within 40ms i.e. very much smaller than the time constants we are trying to measure. This method is shown in Fig.~\ref{fig:LEDmethod}. Note the two transition regions defined by time constants \(\tau_{chargeup}\) and \(\tau_{chargedown}\) shown in this figure. In the charge-down phase of the measurement we can directly observe the detrapping time constant \(\tau_{detrap}\). Since the pixel contains no signal charge, there is nearly no trapping occurring here.  In the charge-up phase we observe some combination of \(\tau_{trap}\) and \(\tau_{detrap}\) since here  both trapping and detrapping currents are present.

\subsection{Measurement of Detrapping time constants.}
\label{sect:expmethod}

\begin{figure}[htbp]
\begin{center}
\begin{tabular}{c}
\includegraphics[height=9cm]{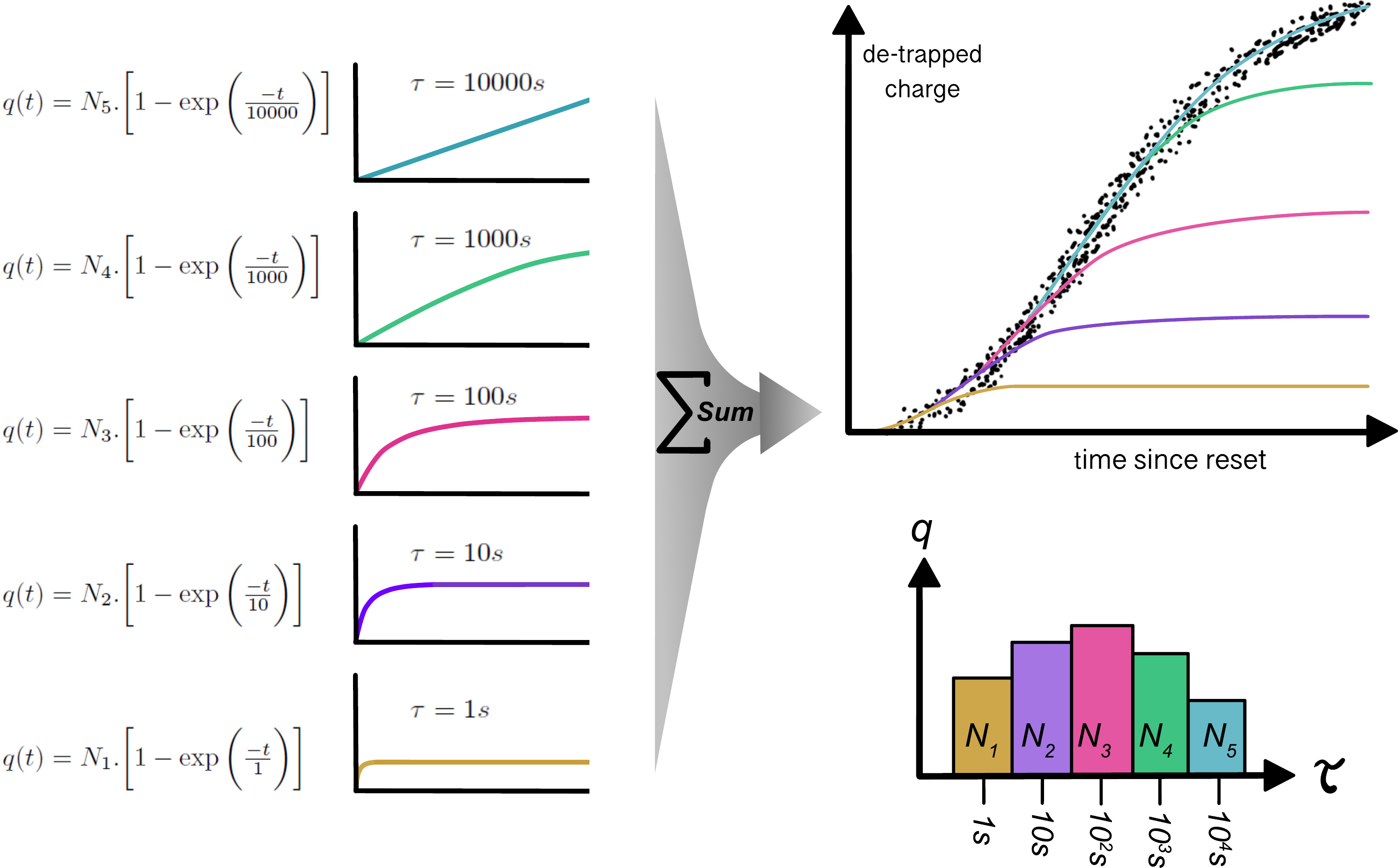}
\end{tabular}
\end{center}
\caption
{ \label{fig:EXPanalysis}
Exponential analysis. The detrapping profile can be analysed as a sum of exponential functions. The top-right panel shows how the 5 functions cumulatively sum to the observed profile.}
\end{figure}

The first measurement made was of the charge-down phase. The LED pulse width was chosen to give an exposure that was a large fraction of the full-well so as to produce enough trapped charge to measure with high signal to noise. The ``soak time'' i.e. the time between LED flash and reset was made long, typically 10000s or longer, so as to ensure that the trapping and detrapping currents had time to equilibrate, even in the presence of traps with long time constants. Following reset, the signal on the pixels was measured non-destructively for 8500s using a series of up to 200 frames spaced over a  geometrically increasing time series. The resultant detrapped charge profile was then analysed as the sum of five exponential functions as illustrated in Fig.~\ref{fig:EXPanalysis}. Each function represented a different time constant bin. In reality the traps in the device will have a continuous distribution of time constants but a Chi-square analysis showed that using more than five bins gave degeneracies in the fit. The chosen bins were 1,10,100,1000 and 10000s. An additional constant offset term was needed in the fit to represent the charge that was already detrapped prior to the first post-reset image being read out, a small delay being necessary to allow the reset transient to decay. The fit was done automatically using the IDL \texttt{MPFIT} function and the user-defined fit function shown in Eq.~\ref{eq:SigmaTau}. The output of the analysis was five numbers \(N_{1..5}\)  representing maximum charge trapped in a pixel for each of the  \(\tau\) bins. Note that this maximum trapped charge will only be obtained at very long soak times. Equation~\ref{eq:SigmaTau} shows the fitting function where \(Q(t)\) is the trapped charge in a pixel as a function of \(t\), the time since reset.
\begin{equation}
Q(t)=\displaystyle\sum_{i=1}^5\bigg[N_i
.\bigg(1-\exp\big({\frac{-t}{\tau_i}\big)}\bigg)\bigg] \\
\label{eq:SigmaTau}
\end{equation}

If the soak time was made sufficiently large these five numbers then represent the equilibrium trapped charge in a pixel for each of the \textit{detrapping} \(\tau\) bins. Rather than do these measurements on a pixel-by-pixel basis, the median signal in a 64x64 pixel window was measured to improve the SNR. The median value was initially measured using a Gaussian fit to the histogram of pixel values in the measurement windows to reject dead and hot pixels. This was later found to be no better than the built-in IDL \texttt{MEDIAN} procedure. Since the signal was so low it was important to  remove the effects of both read-induced glow (i.e. light emission from the transistors in the pixel unit cell), dark current and stray light. This was done using vertically adjacent reference windows (consisting of real pixels \textit{not} reference pixels) that were selectively reset immediately following the LED flash. Due to both dark current and stray light exhibiting considerable spatial variation, additional dark runs with no LED illumination were required to map out these variations. The use of vertically adjacent windows also had the advantage of removing low frequency noise (i.e. baseline drifts) in the signal processing chain as both signal and reference windows would be subject to the same drifts.

\subsection{Measurement of Trapping time constants}
\label{sect:variablesoakmethod}
\begin{figure}[htbp]
\begin{center}
\begin{tabular}{c}
\includegraphics[height=9cm]{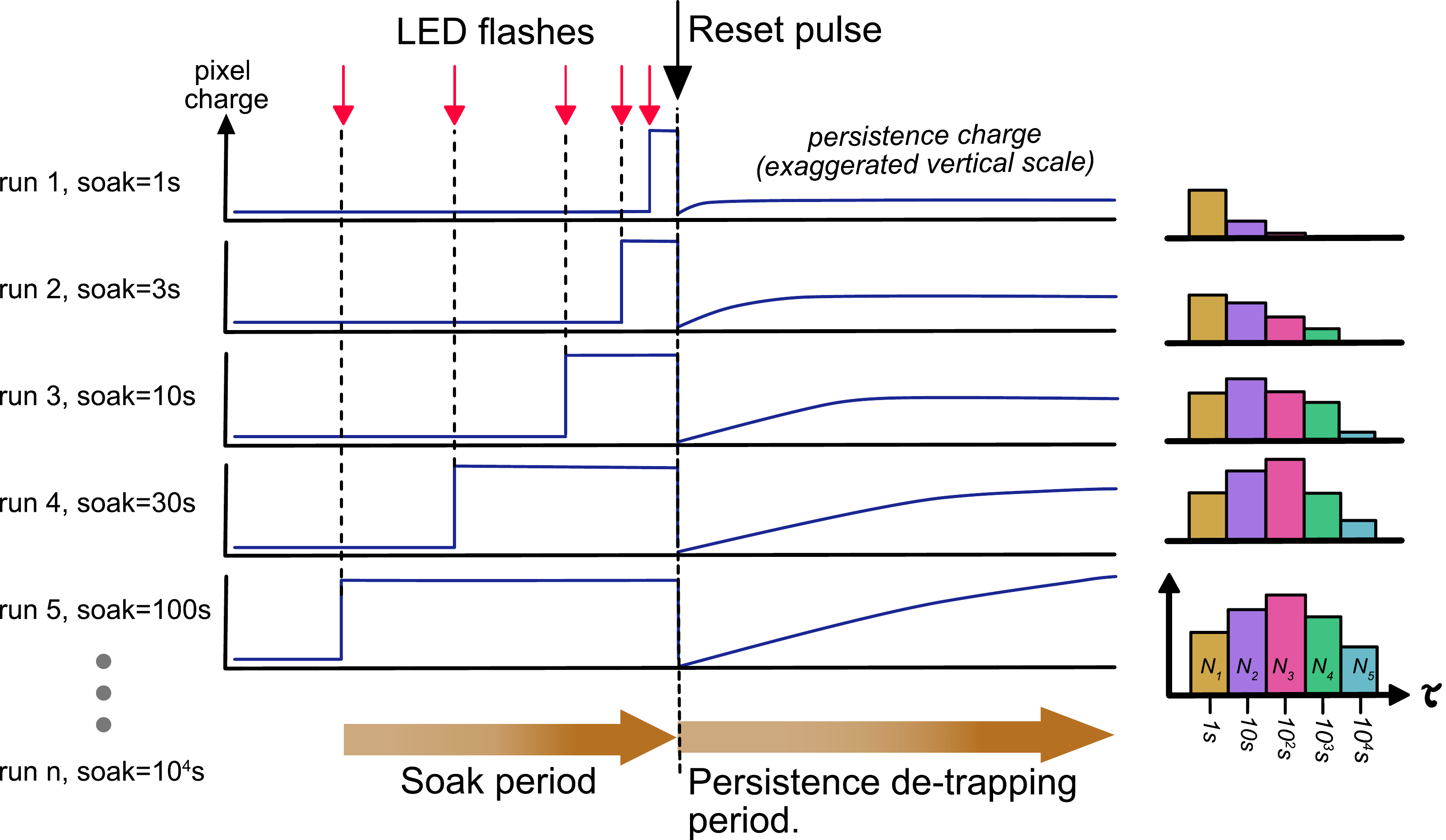}
\end{tabular}
\end{center}
\caption
{ \label{fig:variablesoakmethod}
The charge-up time constants can be characterised by varying the soak time and measuring how the detrapping profile changes in response. This is not real data and for illustration only.}
\end{figure}

Although it is not possible to directly observe the trapping time constants (since trapping current is always accompanied by an opposing detrapping current), they can be inferred by measuring how the detrapping profile changes in response to incremental changes in the soak time. This is illustrated in Fig.~\ref{fig:variablesoakmethod}.

\begin{figure}[htbp]
\begin{center}
\begin{tabular}{c}
\includegraphics[height=9cm]{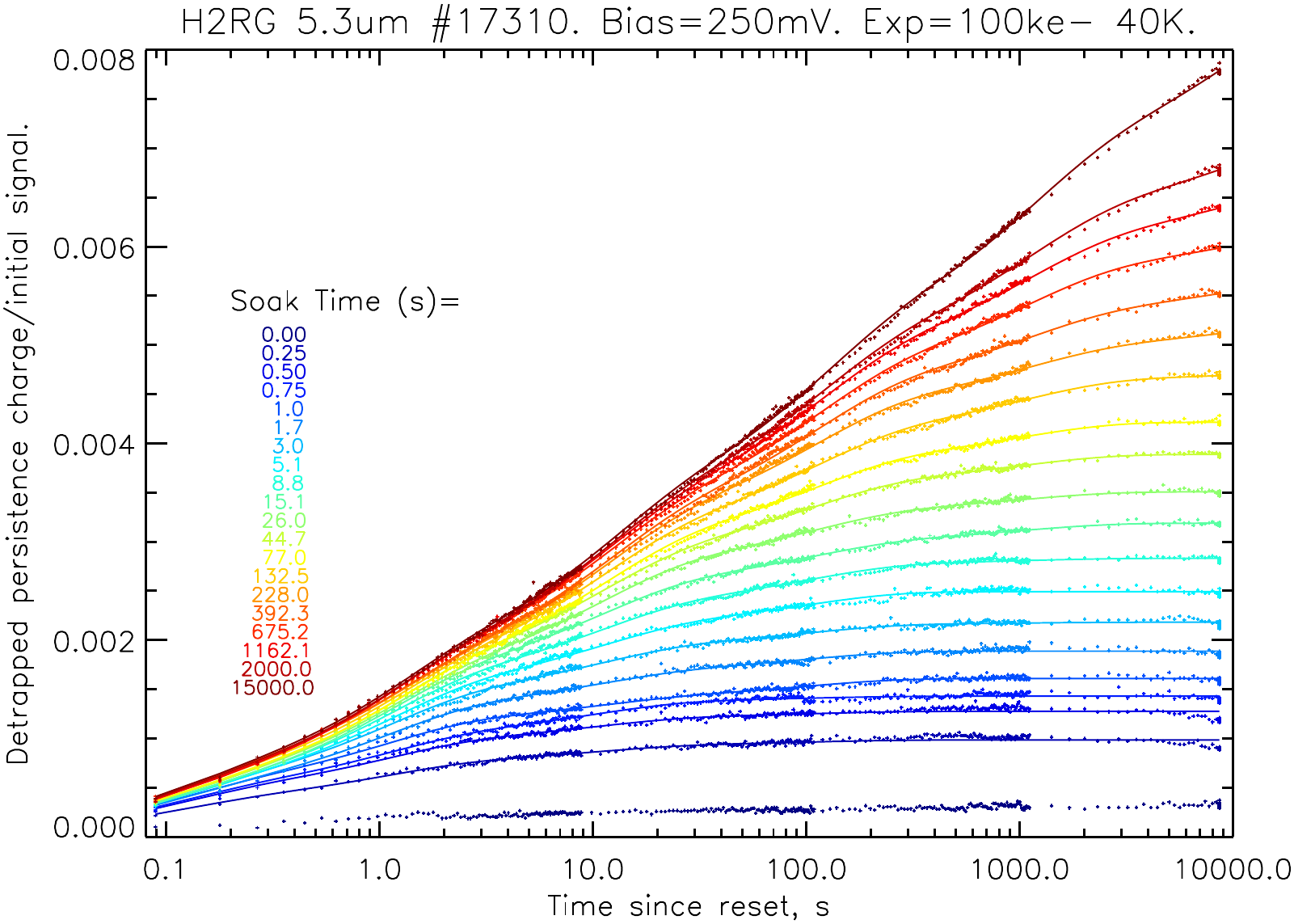}
\end{tabular}
\end{center}
\caption
{ \label{fig:criresPersistversusSoak}
Family of detrapping profiles for the CRIRES+ detector at a constant exposure level of 100ke\(^-\) applied for a soak time that varied between 0 and 15000 seconds. The line fits to the data use Eq.~\ref{eq:SigmaTau}.}
\end{figure}

\begin{figure}[htbp]
\begin{center}
\begin{tabular}{c}
\includegraphics[height=9cm]{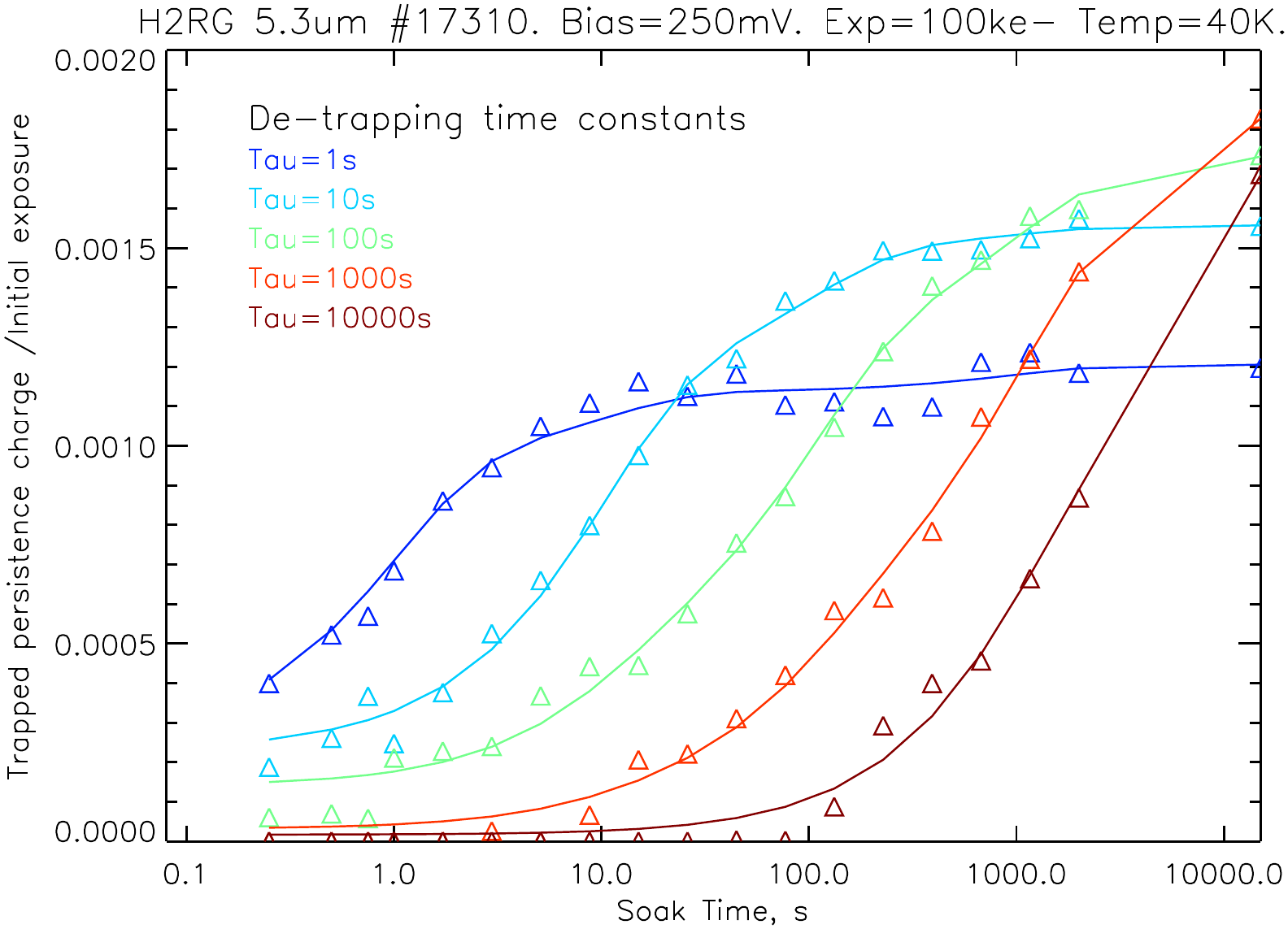}
\end{tabular}
\end{center}
\caption
{ \label{fig:crirestrapgrowth}
The effect of increasing soak time on the amount of charge trapped in each \textit{detrapping} time constant bin. This data is obtained by applying the exponential analysis technique to the data in Fig.~\ref{fig:criresPersistversusSoak}.}
\end{figure}

It was clear from a cursory look at the data that the tendency was for short soak times only to produce short period persistence. Likewise, long period persistence required long soak periods to become significant. A set of actual variable soak data is shown in Fig.~\ref{fig:criresPersistversusSoak}.
Each profile in this plot was subjected to the exponential analysis and the values of \(N_{1..5}\) then replotted as a function of the soak time. This is shown in Fig.~\ref{fig:crirestrapgrowth}.

Here the values of \(N\) have been converted to fractions of the initial exposure level which is a more natural unit. In this figure we can see the exponential charging of the various detrapping time constant bins. Taking the cyan colored curve as an example then we can see something rather interesting. This curve represents how the amount of charge in the \(\tau=10\)s detrapping bin grows as a function of soak time. It appears to have a shape very similar to what we would expect if its charge-up time constant was also equal to 10s.

\begin{figure}[htbp]
\begin{center}
\begin{tabular}{c}
\includegraphics[height=9cm]{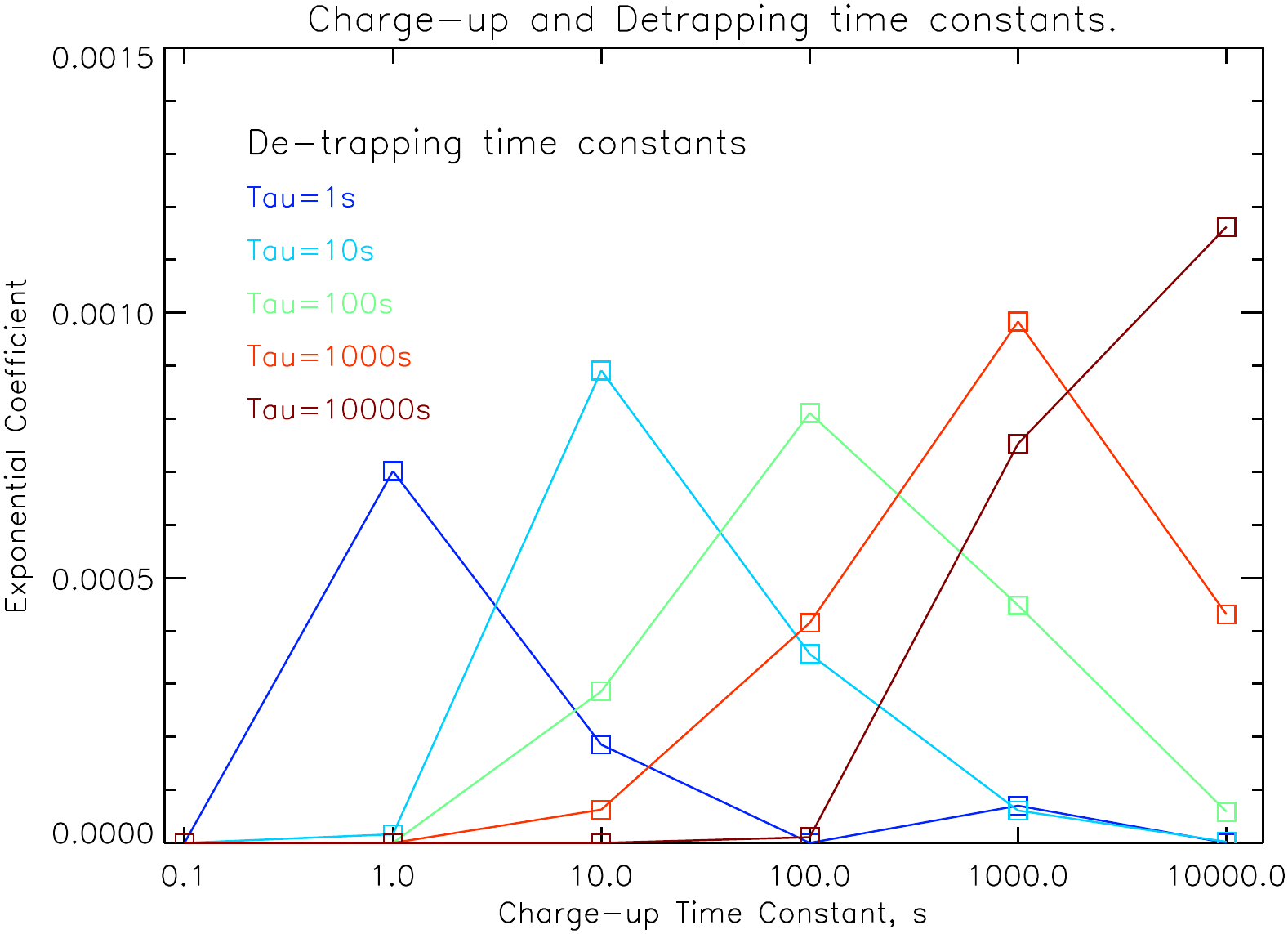}
\end{tabular}
\end{center}
\caption
{ \label{fig:crirestrapdetrap}
Relationship between detrapping and charge-up time constants. This graph finally establishes the link between trap time constants in the post-reset detrapping phase and those in the charge-up soak phase. This data is obtained by applying the exponential analysis technique to the plots in Fig.~\ref{fig:crirestrapgrowth}. Both plots use the same vertical axis units.}
\end{figure}

A more precise analysis then involves applying the same exponential analysis to Fig.~\ref{fig:crirestrapgrowth} that we earlier applied to Fig.~\ref{fig:criresPersistversusSoak}. The result, shown in  Fig.~\ref{fig:crirestrapdetrap}, confirms that \(\tau_{chargeup} = \tau_{detrap}\). This has also been illustrated in the lower panel of Fig.~\ref{fig:LEDmethod}.

The implication of this behaviour is that \(\tau_{trap} >> \tau_{detrap}\). To understand this one must bear in mind that during the charge-up phase the pixel is subject to both trapping and detrapping, whilst during charge-down the pixel is only subject to detrapping current. The fact that both these phases have the same profile over time therefore implies that the trapping current must be close to a constant value. Note that the exponential charging of a capacitor through a resistor will also appear as a constant current at small fractions of its time constant when only a small amount of the total potential charge is present on the capacitor. It follows that the number of charges already trapped have a negligible effect on the trapping current and that only a small fraction of available traps are ever filled.

\subsection{Symbolic Trap Model}
\label{sect:revised}
\begin{figure}[htbp]
\begin{center}
\begin{tabular}{c}
\includegraphics[height=3cm]{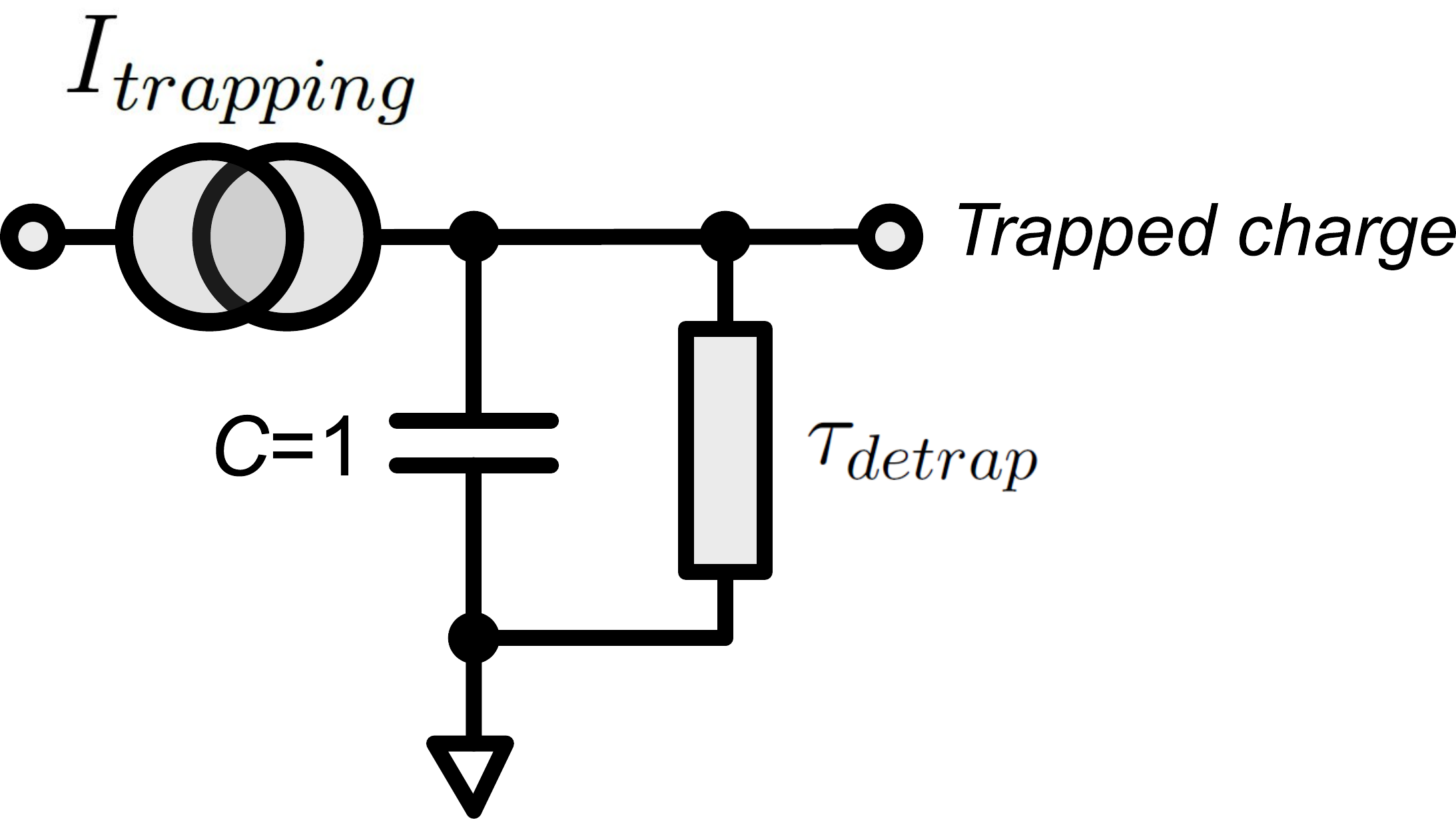}
\end{tabular}
\end{center}
\caption
{ \label{fig:finalmodel}
Symbolic model of an ensemble of persistence-inducing traps. }
\end{figure}

Based on the data presented in the previous section, we can make the following assumptions: 1) detrapping current is proportional to number of filled traps, 2) trapping current is proportional to the number of photoelectrons present in the pixel, and 3) the number of filled traps is equal to the time integral of the difference between trapping and detrapping currents. This allows us to represent the behaviour of the trapping and detrapping currents symbolically with an RC network. This is shown in Fig.~\ref{fig:finalmodel}  for a single time constant bin. The pixel behaviour must be understood as the linear sum of five such bins. The charge on the capacitor corresponds to the trapped photo-charge in a pixel. Charge flows out of (i.e. detraps) the capacitor via a resistor that defines the detrapping time constant. Charge flows in (i.e. traps) via a current source whose value is proportional to the amount of photo-charge available. This symbolic representation of the behavior of the trapping and detrapping currents is helpful in developing the predictive model of the persistence behavior presented later in Sec.~\ref{sect:predict}.


\subsection{ Trap number as a function of signal level.}
\label{sect:probedensityversussignal}
\begin{figure}[htbp]
\begin{center}
\begin{tabular}{c}
\includegraphics[height=10cm]{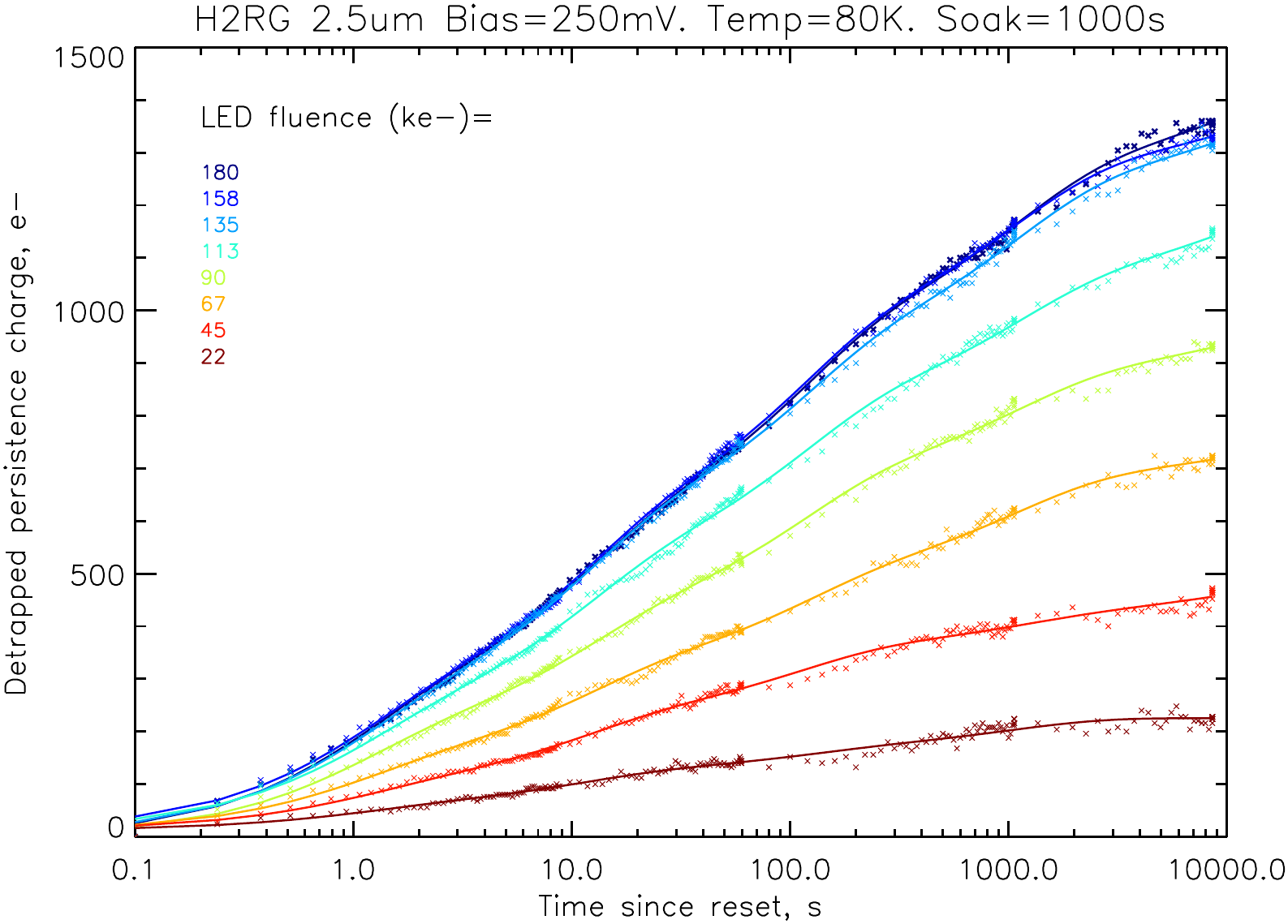}
\end{tabular}
\end{center}
\caption
{ \label{fig:spiffieramplitudes}
Detrapping profiles with constant soak and variable LED fluence for the ERIS 2.5\(\mu\)m detector.  The line fits to the data use Eq.~\ref{eq:SigmaTau}.}
\end{figure}

\begin{figure}[htbp]
\begin{center}
\begin{tabular}{c}
\includegraphics[height=10cm]{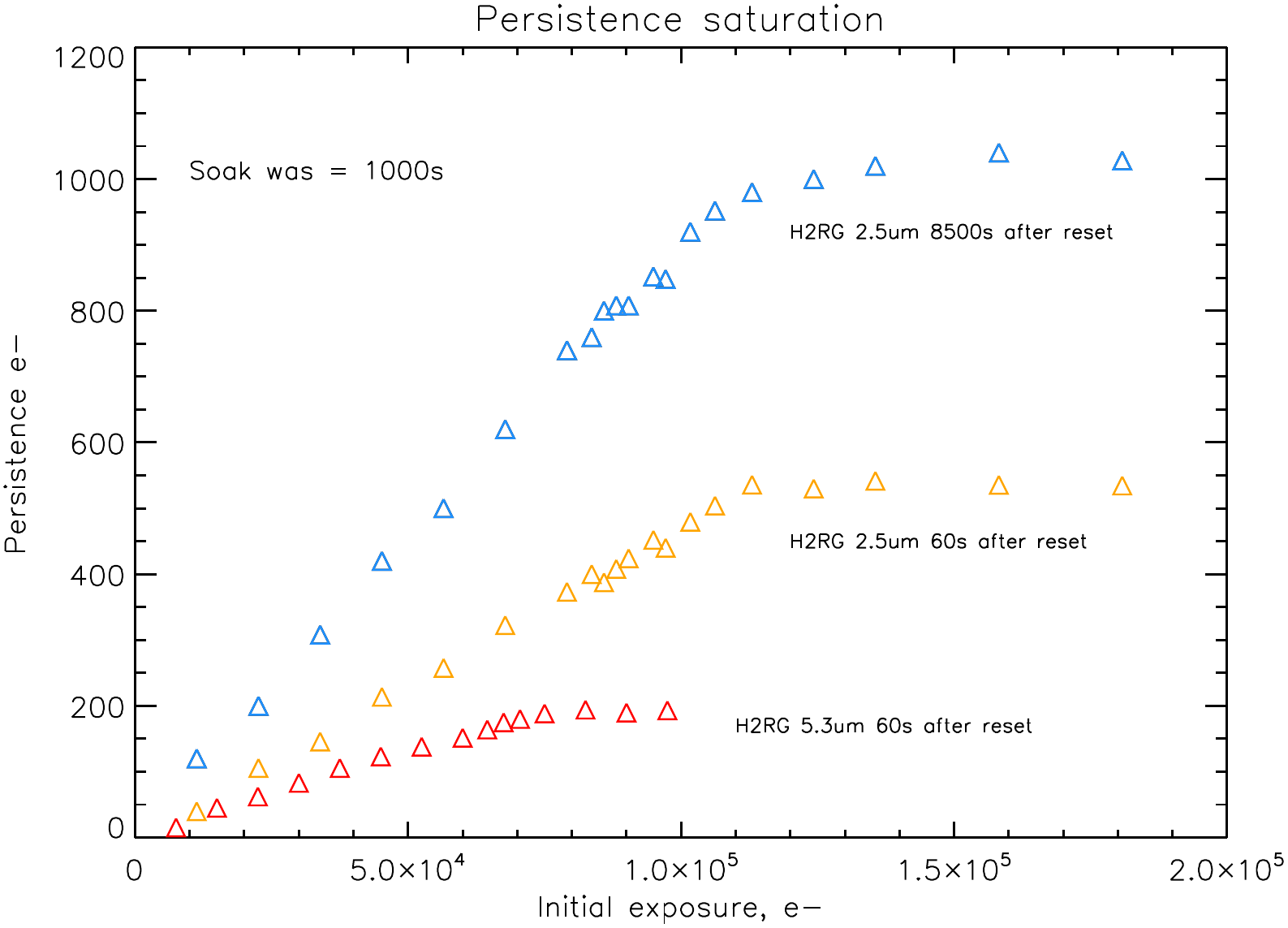}
\end{tabular}
\end{center}
\caption
{ \label{fig:persistenceSaturation}
Persistence saturation at full-well in the ERIS 2.5\(\mu\)m and 5.3\(\mu\)m detectors. Exposure levels were calculated from a combination of LED on-time and its brightness in e\(^-\)/millisecond at low flux levels. Data at 8500s for the 5.3\(\mu\)m device was not included due to the difficulty of accurately removing the high instrumental background contribution.}
\end{figure}

The time constant study used a fairly high illumination level to involve traps throughout most of the depth of the DR and give a persistence signal that could be easily measured. The relationship between trap density, or more precisely, the fraction of the photo-signal stored in traps and initial signal level was then investigated. Soak times were held constant at 1000s and the  LED on-time varied to change the fluence of the exposure. Persistence charge was measured in a small 64x64 window using the technique described at the start of Sec.~\ref{sect:character}. Fig.~\ref{fig:spiffieramplitudes} shows the result of 8 such runs where the exposure level was taken up to and beyond the full-well. With this detector it is clear that for exposures above 135ke\(^-\) the persistence grew no worse. The same data can be represented in another way to show how much charge is detrapped after both 60 and 8500s. Fig.~\ref{fig:persistenceSaturation} clearly shows this persistence saturation effect which is entirely consistent with the depletion region trap hypothesis described in Sec.~\ref{sect:theorytraps}.

Below full-well the relationship between initial exposure (LED fluence) and persistence signal is linear. This greatly simplifies the development of the mathematical model. It is also somewhat surprising since the photodiode is intrinsically quite non-linear in its response to illumination.  Interestingly, Serra et al.\cite{SERRA} have noted a threshold effect in EUCLID-like H2RG detectors where persistence sharply increases above a certain fluence but this was not observed in any of the three detectors used in our study.

\subsection{ Trap time constant variation as a function of signal level.}
\label{sect:probetauversussignal}
\begin{figure}[htbp]
\begin{center}
\begin{tabular}{c}
\includegraphics[height=10cm]{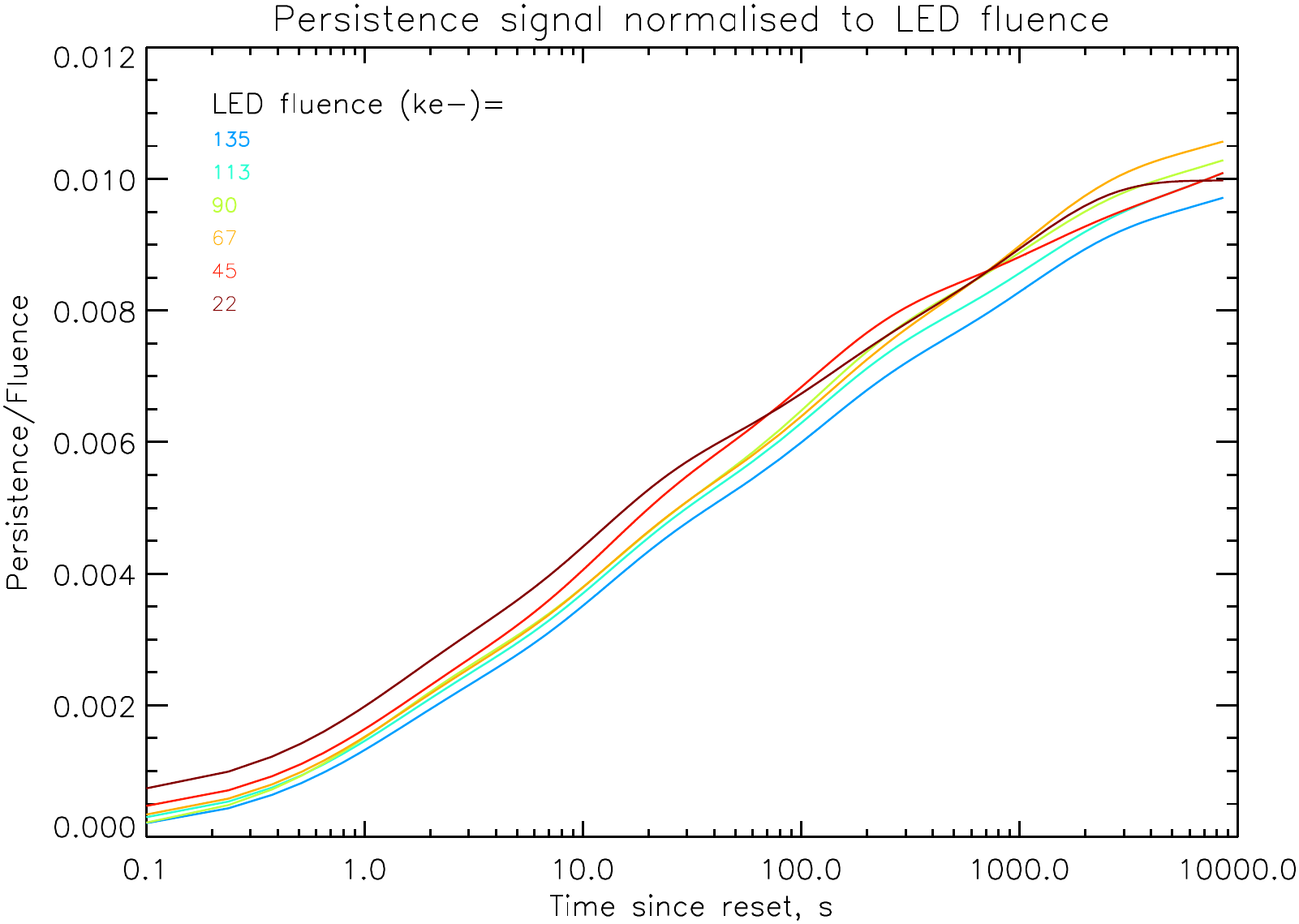}
\end{tabular}
\end{center}
\caption
{ \label{fig:spiffierscaled}
Detrapping time constant invariability with exposure level. The detrapping profiles shown in  Fig.~\ref{fig:spiffieramplitudes} were divided by the initial exposure level to reveal any possible time-constant changes with exposure level.}
\end{figure}

The below full-well data from Fig.~\ref{fig:spiffieramplitudes} was replotted in Fig.~\ref{fig:spiffierscaled} normalised to the LED fluence. This demonstrated that the distribution of trap time constants did not vary throughout the depth of the depletion region i.e. did not vary as a function of signal level. Once again this greatly simplified the mathematical model.

\subsection{ Persistence Maps}
\label{sect:persistencemaps}
\begin{figure}[htbp]
\begin{center}
\begin{tabular}{c}
\includegraphics[height=5.5cm]{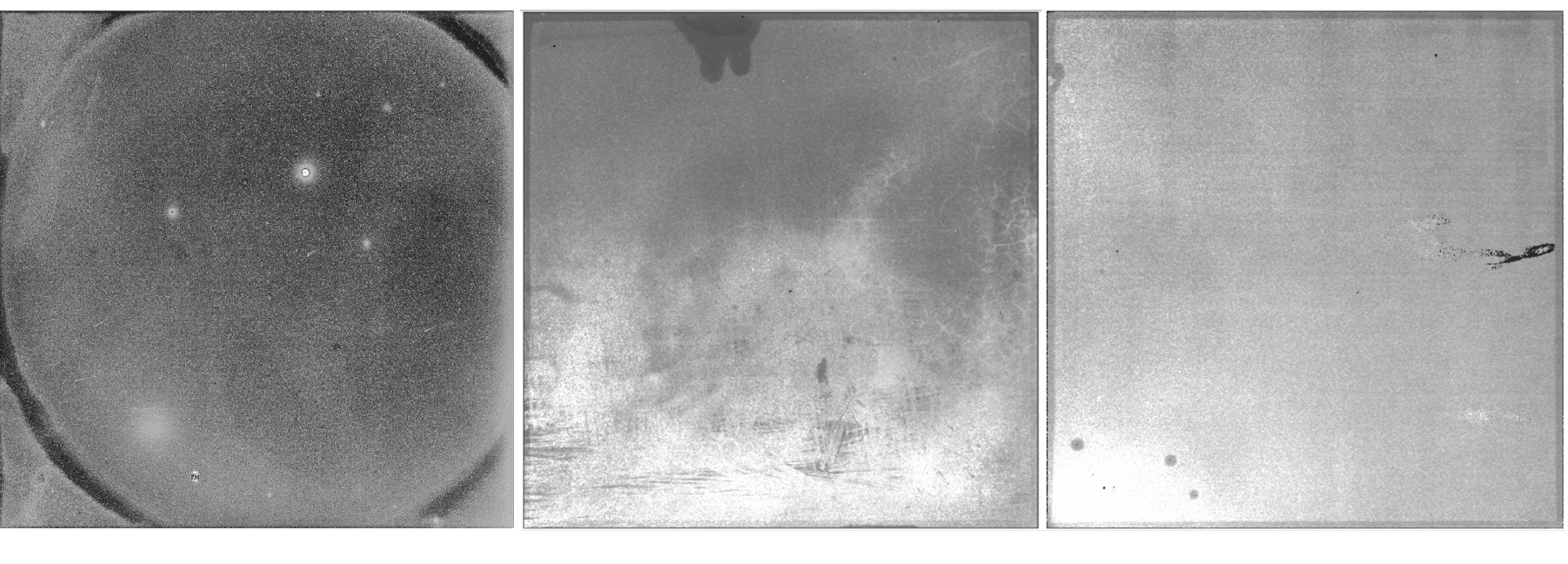}
\end{tabular}
\end{center}
\caption
{ \label{fig:ThreePersistMaps}
Persistence maps for (from left to right) the CRIRES+ 5.3\(\mu\)m, ERIS 2.5\(\mu\)m and 5.3\(\mu\)m detectors. }
\end{figure}

Spatial variations in trap number per pixel were mapped out by averaging long sequences of 60s dark frames that followed an initial deep LED exposure followed by a 60s soak. The use of relatively short exposure times reduced the dark current contribution. The sequences were nonetheless repeated but with the LED disconnected to produce dark reference maps that were later subtracted from the persistence+dark maps. It was assumed as a starting point that the distribution of trap time constants did not vary over the detector, and the only thing that varied from pixel to pixel was the total number of traps.  Three maps are shown in Fig.~\ref{fig:ThreePersistMaps}. The ERIS 2.5\(\mu\)m detector had an anomalous region towards the top where the persistence was about 1/4 of the average. This region was associated with a glue void between multiplexer and photodiode material. A glue void on the CRIRES+ detector, however, is barely noticeable in its persistence map.

\subsection{Photon-transfer statistics of persistence charge}
\label{sect:PTstats}

It was shown in Fig.~\ref{fig:depletionModel} that electrons moving to and from traps within the depletion region should demonstrate an effective fractional charge i.e. a different e\(^-\) /ADU conversion gain. This gain is typically measured using the Photon Transfer\cite{Janesick} technique that compares the mean and variance of signal charge.  The technique requires image pairs to be taken so as to eliminate fixed pattern noise and ensure that the measured variance was due to Poissonian noise in the actual signal.

In the normal application of the Photon Transfer technique a series of flat field pairs of incrementing exposure depth are used. Here we instead use two series of detrapped-charge images taken in quick succession. Each series consists of a deep LED illumination, a 60s soak followed by a reset and a few dozen non-destructive reads to capture the steadily increasing detrapped charge over the course of a few minutes. A 64x64 window aligned to a single H2RG amplifier was used. The measurements were extremely challenging due to the very low signal levels. Higher signal levels were obtained by using longer soak times but this of course increased the interval separating the two series and allowed in too much 1/f noise. Consequently even when using reference pixel subtraction, no meaningful statistical analysis was possible.

\begin{figure}[htbp]
\begin{center}
\begin{tabular}{c}
\includegraphics[height=7cm]{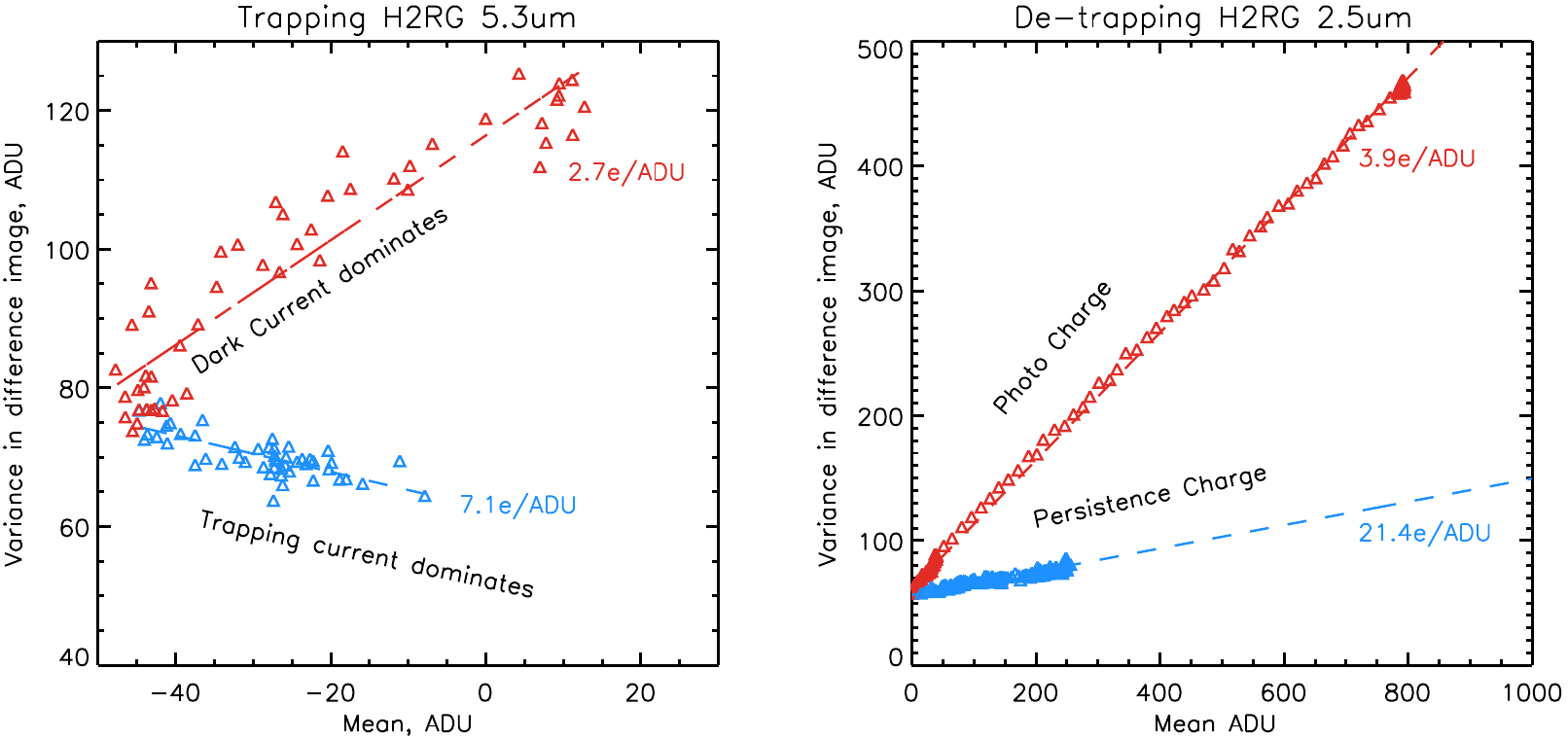}
\end{tabular}
\end{center}
\caption
{ \label{fig:persistPT}
Photon transfer of persistence. The left panel (ERIS 5.3\(\mu\)m device) illustrates the statistics of the trapping current in blue and the dark current in red. The negative gradient of the trapping current statistics is due to the mean signal on the pixel actually decreasing due to the trapping. Dark current of course acts to increase the signal. The panel on the right (ERIS 2.5\(\mu\)m device) shows the statistics of detrapping current in blue and photo-charge/dark current in red. All of these act to increase the signal on the pixel.}
\end{figure}

The gain of both trapped and detrapped charge was measured. The detrapped charge, as shown in the right panel of Fig.~\ref{fig:persistPT}, had a variance that was 20\% of that measured from photoelectrons. Measuring the statistics of the trapped charge was done slightly differently by analysing the statistics of the portion of the up-step profile (see Fig.~\ref{fig:LEDmethod}) lost to traps immediately following the LED exposure. Unfortunately this method could only be applied to the ERIS 5.3\(\mu\)m device due to time limitations and since this detector suffered from stray light, the quality of the data was poor. For the first few tens of seconds following the LED flash the signal on the pixel was seen to decrease due to the dominant trapping current but rapidly reached an inflexion point as the stray light then caused the signal to rise once again. The result is shown in the left panel of Fig.~\ref{fig:persistPT}. Here the first 50 data points corresponding to image pairs taken shortly after the start of illumination are color coded blue whilst those taken at later times coded red. The blue data shows predominantly the statistics of the charge trapping whilst the red data shows that of the dark current. The sub-Poissonian nature (i.e. lower-than-expected variance) of the charge trapping is demonstrated as can be seen from the differing gradients of the linear fits to each set of data.
\subsection{Role of charge trapping in Reciprocity Failure}
\label{sect:reciproc}
\begin{figure}[htbp]
\begin{center}
\begin{tabular}{c}
\includegraphics[height=8cm]{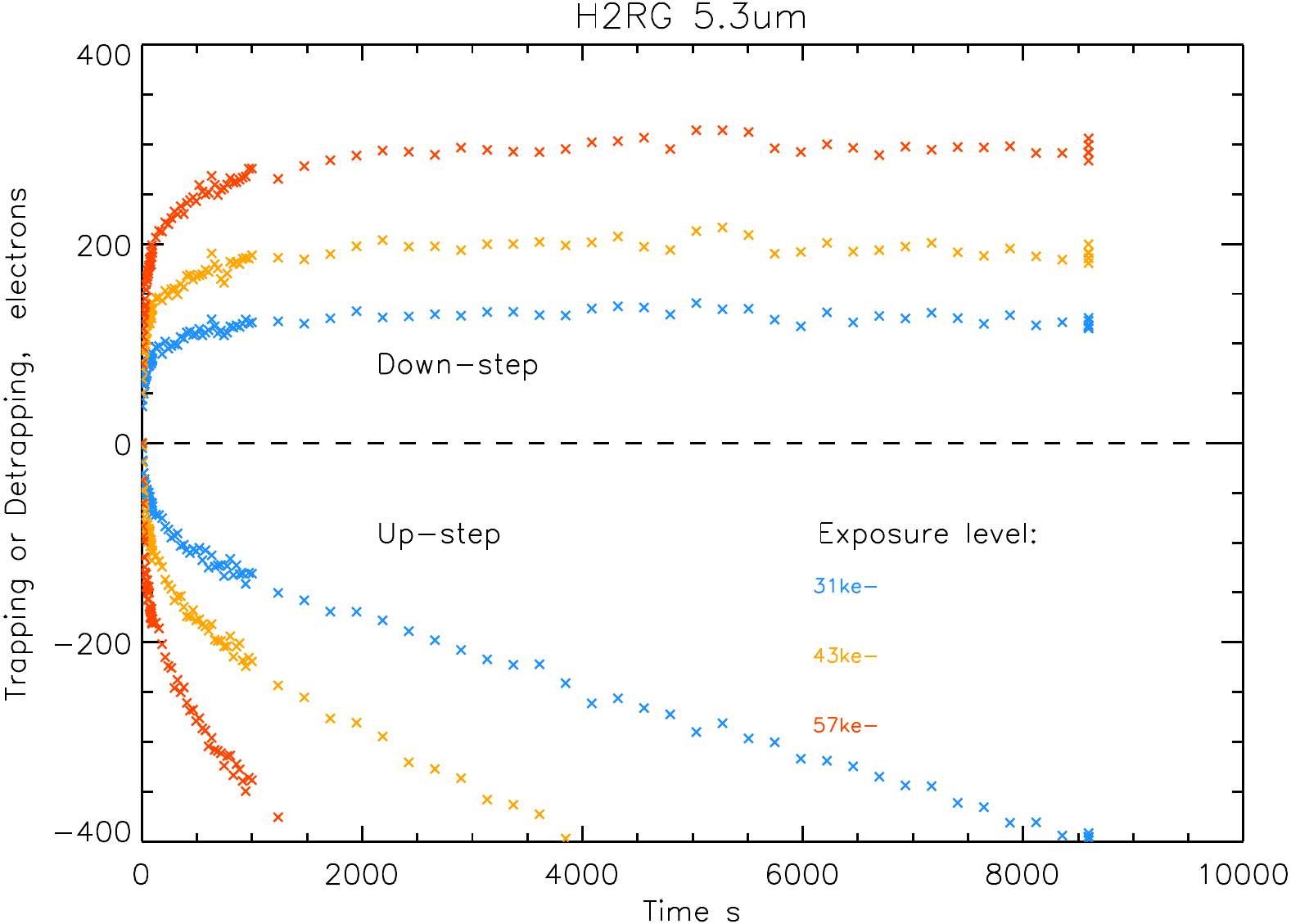}
\end{tabular}
\end{center}
\caption
{ \label{fig:mirror1}
Charge loss from a pixel due to trapping currents during the ``up-step'' exposure phase and charge gain due to detrapping currents during the ``down-step''. The up-step runs reveal an additional leakage current in the pixel. Refer to Fig.~\ref{fig:LEDmethod} for an explanation of the terms. Data obtained using the ERIS detector.}
\end{figure}

\begin{figure}[htbp]
\begin{center}
\begin{tabular}{c}
\includegraphics[height=8cm]{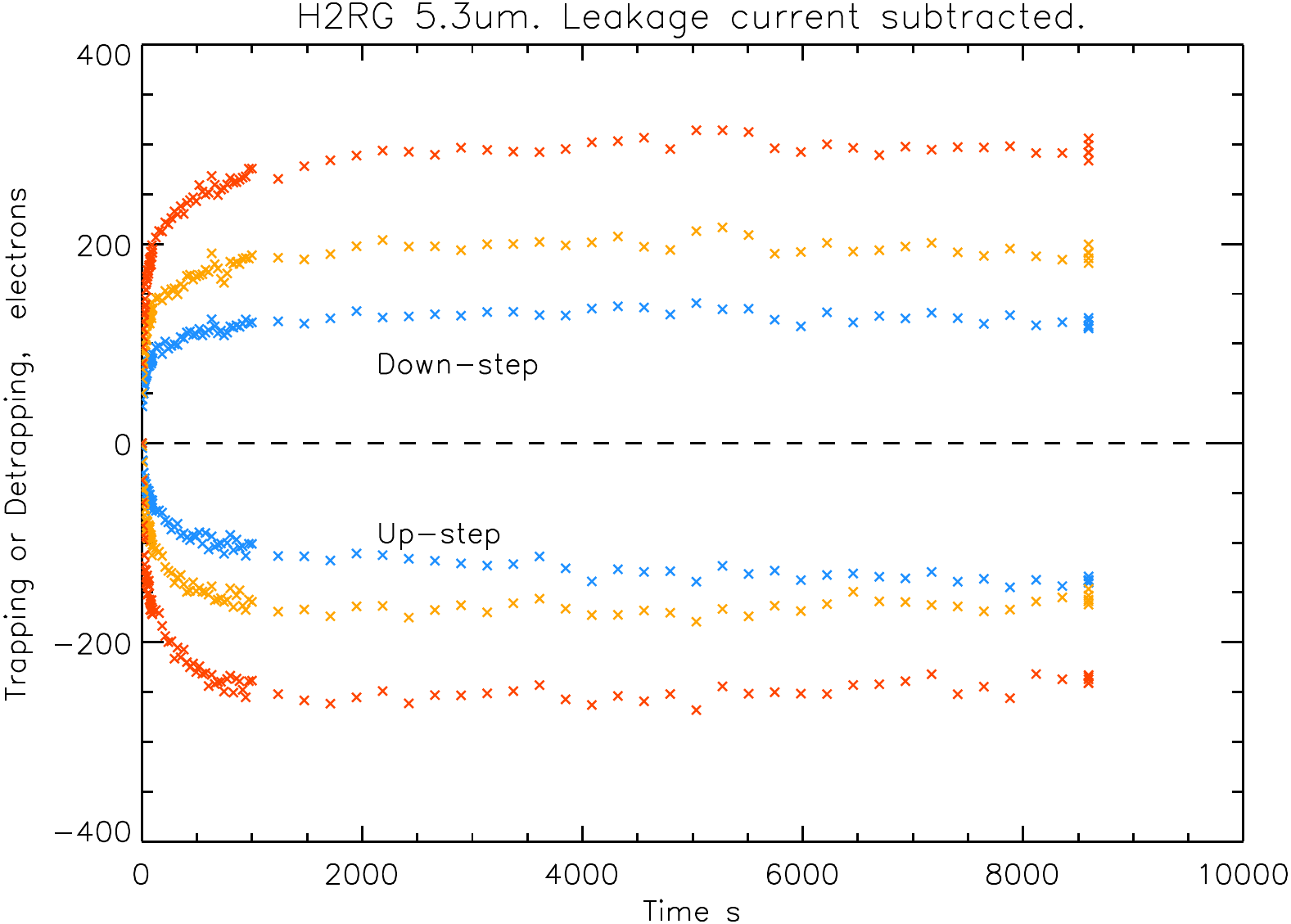}
\end{tabular}
\end{center}
\caption
{ \label{fig:mirror3}
Symmetry of ``up-step'' exposure phase and charge gain due to detrapping currents during the ``down-step'', as shown in Fig.~\ref{fig:mirror1}, once the leakage current has been subtracted. Refer to Fig.~\ref{fig:mirror1} for the color codes of the three plot groups.}
\end{figure}

The precise shape of the up-step profile (i.e. the signal loss due to trapping current exceeding detrapping current during the soak period) should mirror that of the down-step profile given the earlier conclusion that the charge-up time constant is equal to the detrapping time constant. Fig.~\ref{fig:mirror1} shows one such measurement run analysed in three separate windows each with a different illumination level. The symmetry of the up-step and down-step profiles is not perfect. It seems that the exposed pixel, although dominated by trapping current at short times, becomes dominated by a secondary leakage current of unknown origin at longer times. This leakage current becomes more serious as full-well is approached and can be many times the dark current although of opposite sign. At a signal level of 66ke\(^-\) it was measured at 0.24e\(^-\)/s in the ERIS 5.3\(\mu\)m device. At the low signal levels encountered when measuring the detrapping profiles the leakage is negligible since these profiles eventually reach zero gradient. We can therefore be confident that our persistence data has not been degraded by leakage. As a final check of this, the up-step data in Fig.~\ref{fig:mirror1}  was replotted with the leakage currents (assumed to be constant with time but varying with initial exposure level) subtracted. The result is shown in Fig.~\ref{fig:mirror3}. The symmetry between the up and down-step phases is nicely demonstrated.

This charge loss, both due to trapping and leakage current, must be a major contributor to Reciprocity Failure in H2RG detectors. Regan et al.\cite{Regan} also reached the conclusion that traps are at least a partial explanation. Biesiadzinski et al. \cite{Biesiadzinski} have measured reciprocity failure using photometric techniques. They discuss leakage currents and trapping as possible explanations. Interestingly they also note that low operating temperatures reduce Reciprocity Failure which further supports the trap explanation.

\subsection{Trap analysis results for all three detectors}
\label{sect:results}
The characterization of the ERIS 5.3\(\mu\)m detector has already been presented in Fig.~\ref{fig:criresPersistversusSoak}. Similar plots were obtained for the other two detectors. In all three plots the detrap profile of the longest soak time used was then subjected to the exponential analysis to give the trap densities at the 5 time constants. The results are shown in Table~\ref{tab:Taus}.

\begin{table}[ht]
\caption{Trap analysis results. The data shows  \(\rho\): the maximum fraction of the photo-charge that is trapped in each time constant bin at equilibrium.}
\label{tab:Taus}
\begin{center}
\begin{tabular}{|l|l|l|l|} 
\hline
\rule[-1ex]{0pt}{3.5ex} \(\tau\)  & CRIRES+ 5.3\(\mu\)m \#17310 & ERIS 2.5\(\mu\)m \#G18832 & ERIS 5.3\(\mu\)m  \#G18853 \\
\hline\hline
\hline
\rule[-1ex]{0pt}{3.5ex}  1s & 1.4e-3 & 1.6e-3 & 5.9e-4  \\
\hline
\rule[-1ex]{0pt}{3.5ex}  10s & 1.5e-3 & 3.2e-3 & 1.7e-3   \\
\hline
\rule[-1ex]{0pt}{3.5ex}  100s & 1.7e-3 & 3.9e-3 & 1.5e-3 \\
\hline
\rule[-1ex]{0pt}{3.5ex}  1000s & 1.8e-3 & 4.9e-3 & 1.6e-3  \\
\hline
\rule[-1ex]{0pt}{3.5ex}  10000s & 1.7e-3 & 5.0e-3 & 1.3e-3 \\
\hline
\end{tabular}
\end{center}
\end{table}

\section{Predictive Model Development}
\label{sect:predict}  
The predictive model of persistence needs to keep a continuous ledger of the charge trapped in each time constant bin. If this quantity decreases during our current exposure then it will contribute to our image as persistence. Conversely if it increases then our current exposure will suffer charge loss. The input to the model is a continuous stream of image data. This data should be as complete as possible i.e. there must be no unrecorded exposures. Any gaps in the recorded exposure history of our detector will reduce the effectiveness of the model.

In Sec.~\ref{sect:character}, we showed that our data supports three assumptions that allow the trapping and detrapping currents to behave in the same way as the symbolic model shown in Fig.~\ref{fig:finalmodel}, therefore a mathematical description of this symbolic model will also describe the behavior of the trapped charge.
As a starting point the charge on the capacitor (i.e. the charge in each trap time constant bin) can be expressed as the integral over time of the balance between trapping current through the current source and de-trapping current through the parallel resistor. This is given by:
\begin{equation}
Q(t)=\int_0^t[I_{trap}(t)-I_{detrap}(t)] dt.
\label{eq:Equation1}
 \end{equation}
The de-trapping current is proportional to the charge that has already been trapped and is described by:
\begin{equation}
I_{detrap}(t)=\frac{Q(t)}{\tau}.
\label{eq:Equation2}
\end{equation}
As shown in Sec.~\ref{sect:revised} the trapping current is independent of the already-trapped charge and approximates to a current source proportional to the exposure level:
\begin{equation}
I_{trap}(t)=\frac{\rho(\tau).E(t)}{\tau},
\label{eq:Equation3}
\end{equation}
where \(E(t)\) is the photo-charge on the pixel and \(\rho(\tau)\) the fraction of trapped photo-charge at equilibrium. The independence from already-trapped charge can be understood as being a consequence of the  long trapping time constants. The charge flowing into a capacitor at times very short compared to its RC constant will approximate very closely to a constant current i.e. will be independent of how much charge is already stored.

Combining Eqs.~\ref{eq:Equation1},~\ref{eq:Equation2} and ~\ref{eq:Equation3} and summing over the 5 time constant bins we obtain:
\begin{equation}
Q(t)=\displaystyle\sum_{i=1}^n\displaystyle\int_0^t\bigg[\frac{\rho(\tau_i).E(t)}{\tau_i}-\frac{Q(t-dt)}{\tau_i}\bigg]dt.
\label{eq:Equation4}
\end{equation}

\begin{figure}[htbp]
\begin{center}
\begin{tabular}{c}
\includegraphics[height=8cm]{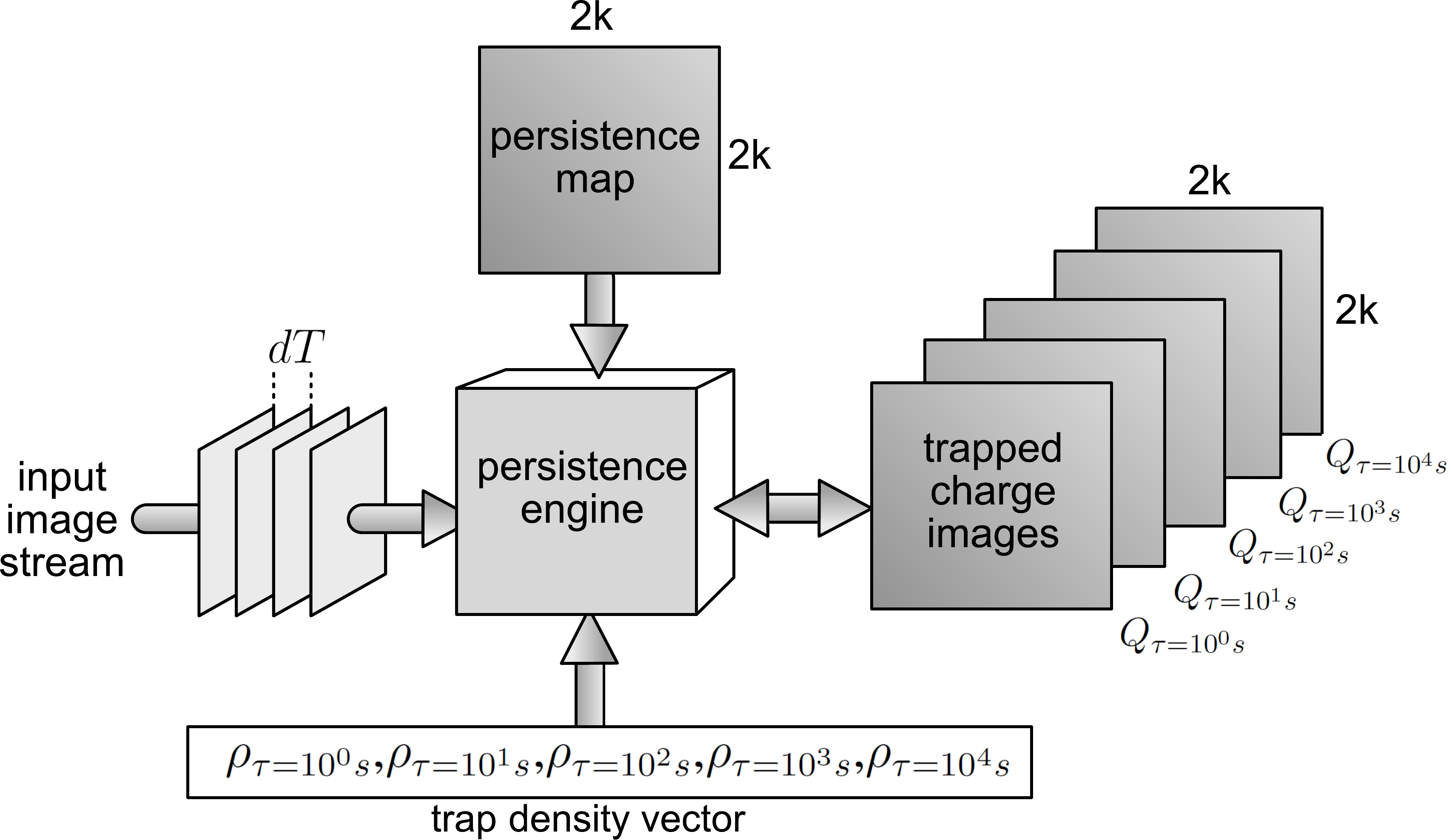}
\end{tabular}
\end{center}
\caption
{ \label{fig:processor}
Main elements of the persistence processor system for the H2RG. }
\end{figure}

\begin{figure}[htbp]
\begin{center}
\begin{tabular}{c}
\includegraphics[height=3cm]{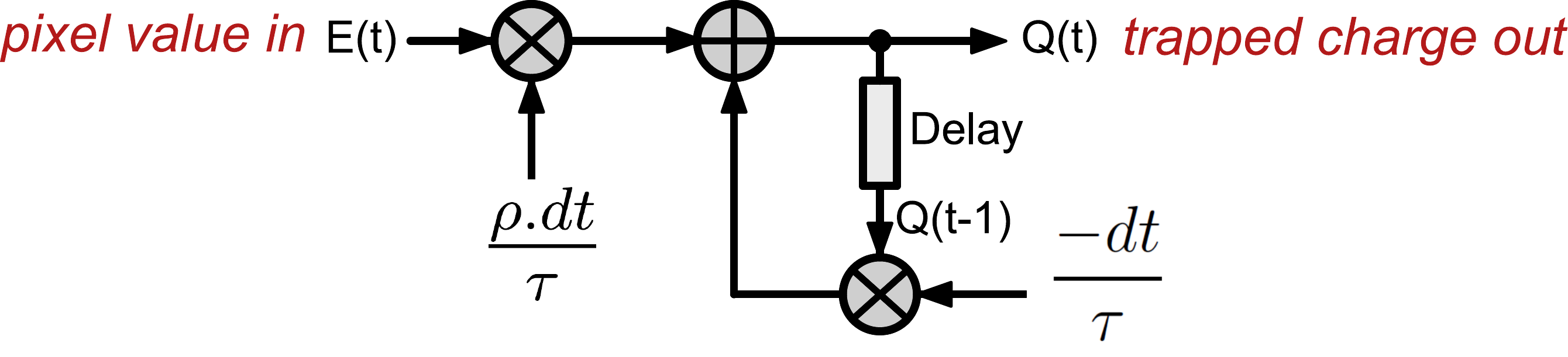}
\end{tabular}
\end{center}
\caption
{ \label{fig:IIR}
The persistence correction algorithm represented as an Infinite Impulse Response digital filter. Five such filters are required, one for each of the time constant bins.}
\end{figure}

This function can then be numerically integrated with \(dt\)~=~time between frames, to yield \(Q(t)\). This process looks exactly like an Infinite Impulse Response (IIR) digital filter\cite{Lyons}. This is a computationally efficient recursive filter which calculates the actual output by combining weighted versions of the actual input (the pixel signal) and the previous output (previous level of trapped charge). It is commonly used in digital signal processing to describe the behaviour of RC networks where any input disturbance to the system produces an output signal with an exponentially decreasing amplitude. The elements of the ``persistence processor'' are  the five IIR filters, a read-only persistence map, 5 read/write trapped charge images and a 5-element trap density vector. The vector does not indicate true densities. Instead it contains dimensionless elements that show the maximum fraction of the pixel photo-charge that can be trapped.
Processor operation is shown schematically in Fig.~\ref{fig:processor}. Here we assume that a single map is sufficient i.e the distribution of trap time constants does not vary between pixels. The trap density vector for a given detector will only be valid for a given operating temperature and bias-voltage setting. It is assumed that it will otherwise remain static, however, long period tests of this assumption have not yet been done.
 Using standard digital filter representation our persistence processor can be described by Fig.~\ref{fig:IIR}. A non-optimised IDL implementation of five such filters (one for each time constant), applied to each pixel in an H2RG image executed in 250ms on a 3GHz desktop using floating point arithmetic. An important requirement of our instruments is that it take less time to reduce the data on an average PC than to acquire it, so any algorithm we introduce must meet that requirement.

\section{Model Accuracy}
\label{sect:accuracy}  
\begin{figure}[htbp]
\begin{center}
\begin{tabular}{c}
\includegraphics[height=6cm]{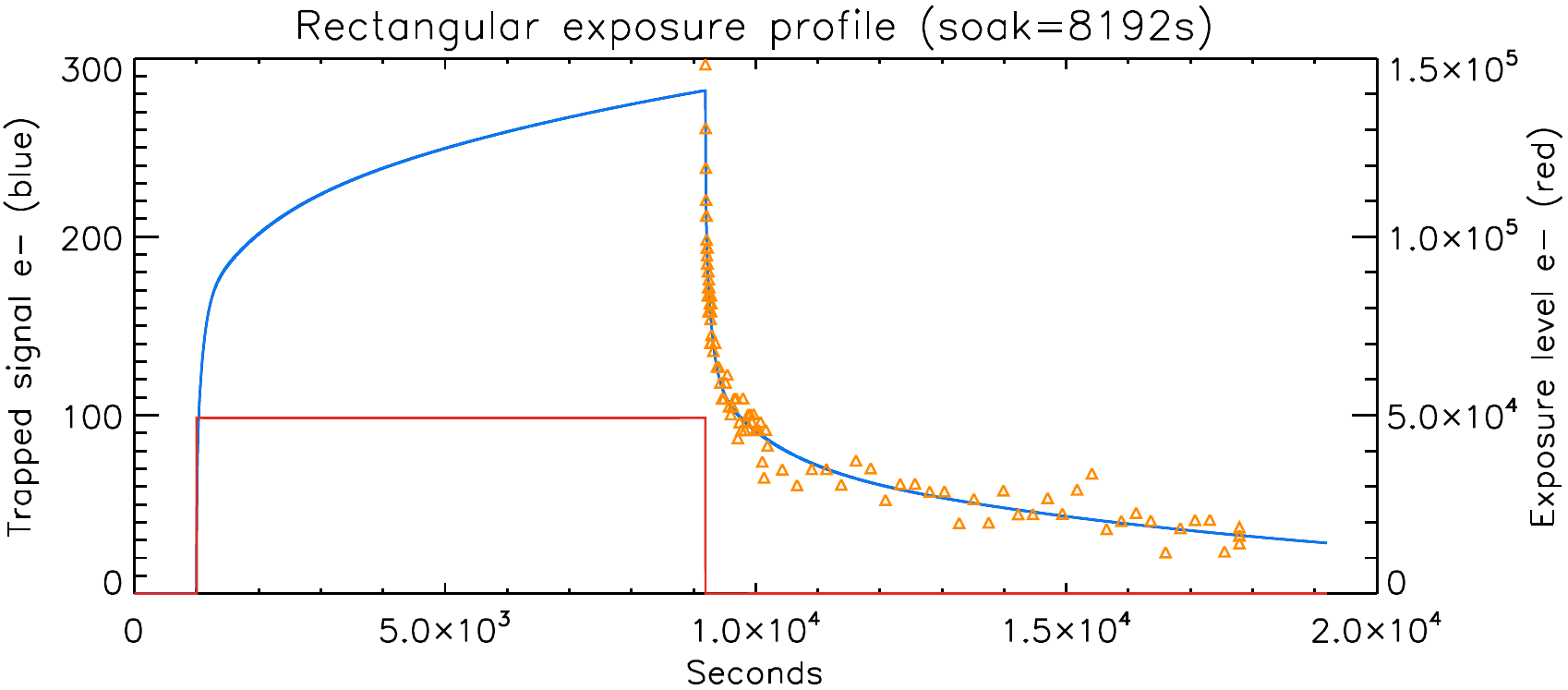}
\end{tabular}
\end{center}
\caption
{ \label{fig:compareplot1}
Comparison of the persistence model with an actual 8192s soak at an exposure level of 50ke\(^-\). Data taken with ERIS 5.3\(\mu\)m detector. The blue plot represents the predicted trapped charge, red the exposure level and the orange triangles the actual data.}
\end{figure}

Initial tests of the model were made using certain simplifying assumptions. The trap density vector elements \(\rho_{1..5}\) were determined for all pixels by simply multiplying the mean values obtained from a 64 x 64 window  by the persistence map. The map had first been normalised so that its average value within the window was 1. The persistence map was the result of 60s soaks and actually contained little information on long period traps. Nevertheless this simplification worked quite well which further supported our initial assumption that the distribution of trap time constants does not vary between pixels, rather it is the total number of traps that varies.
The first test was of a long soak following a medium fluence LED flash. This highlighted an important problem. The frame time of the test was 1.38s which exceeds the minimum trap time constant in the model. Physically this would mean that the 1s traps are almost fully charged within 1 frame time. Unfortunately this causes the filter to oscillate. For stability the frame time \(dt\) must be less than the minimum trap time constant under consideration. The solution here was simply to remove the \(\tau=1s\) filter since it would only affect the first frame after reset anyway. The result is shown in Fig.~\ref{fig:compareplot1}. There is a good fit between the model and the data.

\begin{figure}[htbp]
\begin{center}
\begin{tabular}{c}
\includegraphics[height=6cm]{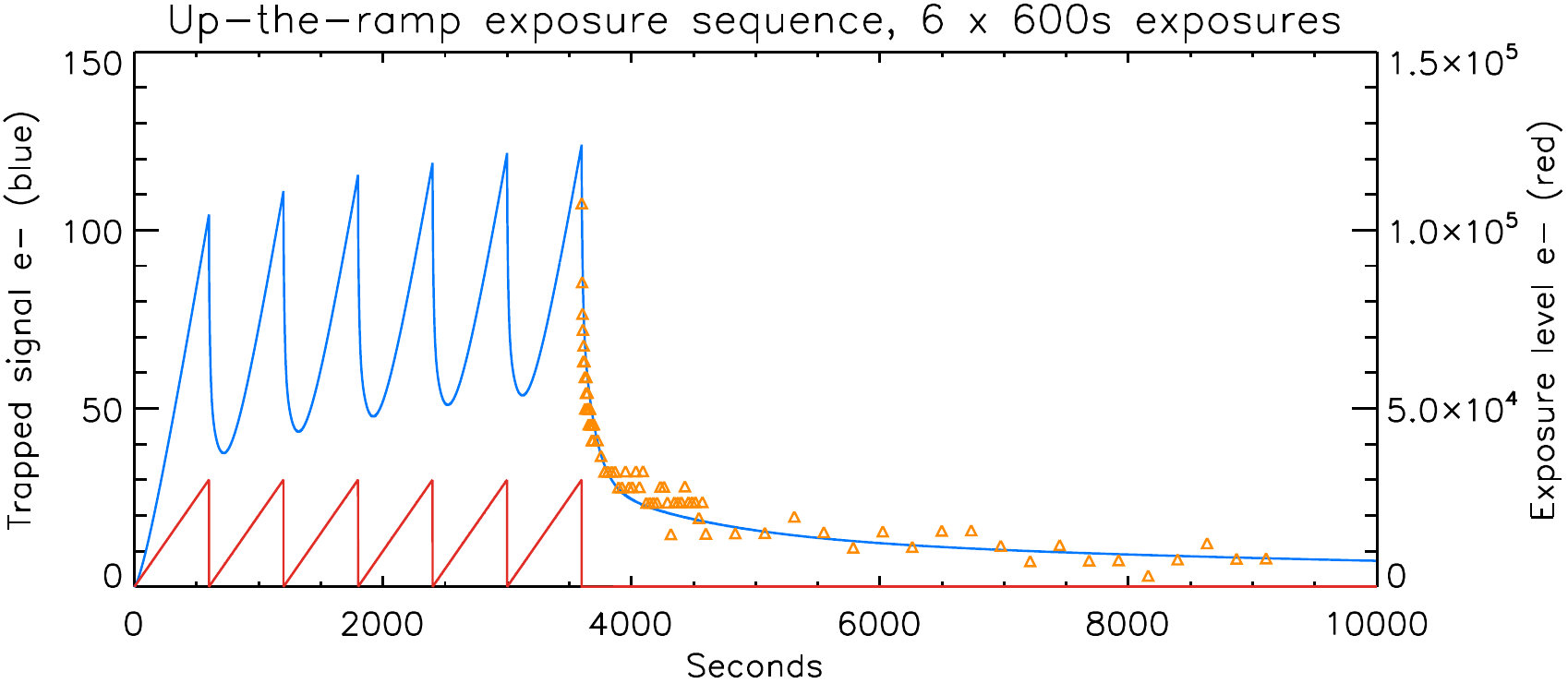}
\end{tabular}
\end{center}
\caption
{ \label{fig:compareplot3}
Comparison of the persistence model with an actual series of 6 up-the-ramp exposures of 600s to a level of 30ke\(^-\). Data taken with ERIS 5.3\(\mu\)m detector. The blue plot represents the predicted trapped charge, red the exposure level and the orange triangles the actual data.}
\end{figure}

\begin{figure}[htbp]
\begin{center}
\begin{tabular}{c}
\includegraphics[height=6cm]{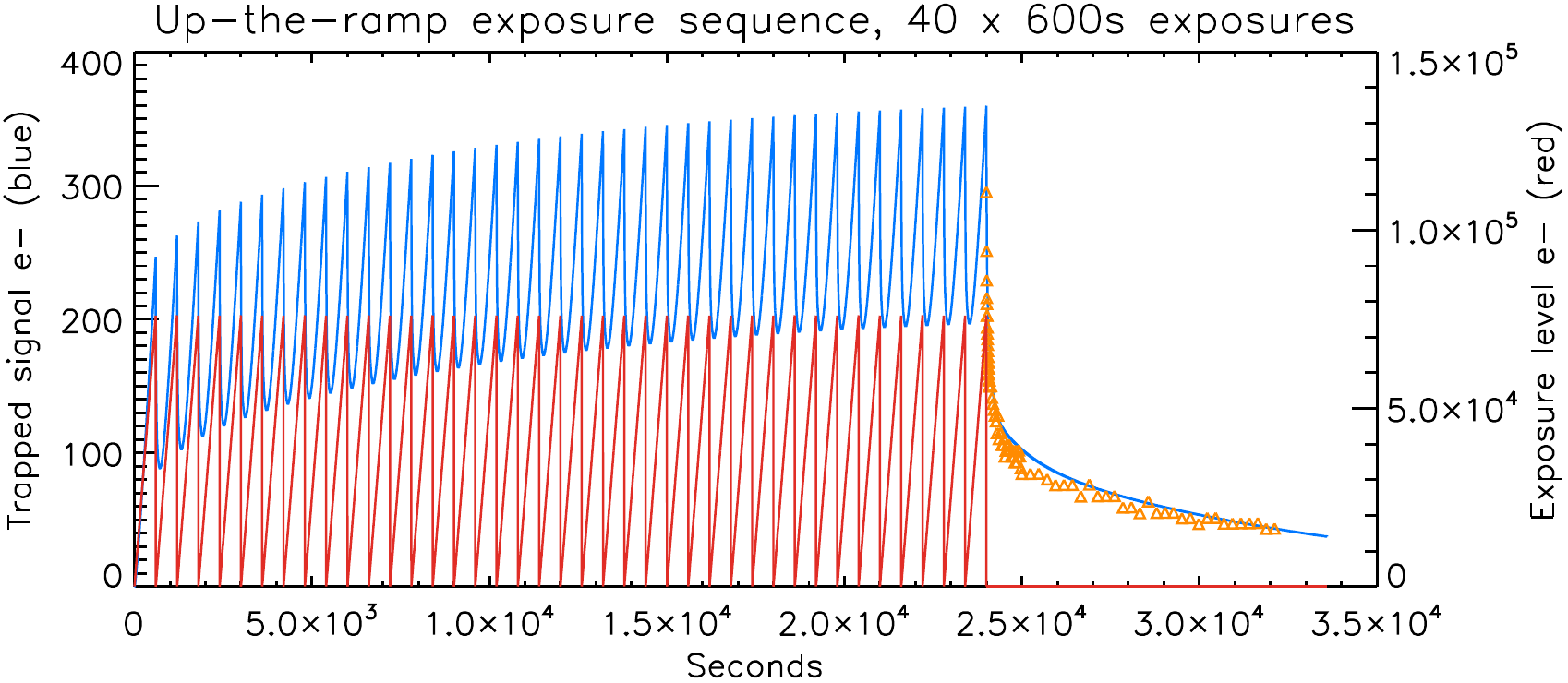}
\end{tabular}
\end{center}
\caption
{ \label{fig:compareplot2}
Comparison of the persistence model with an actual series of 40 up-the-ramp exposures of 600s to a level of 75ke\(^-\). Data taken with ERIS 5.3\(\mu\)m detector. This test probed the model behaviour in a regime where very long time constant traps contained significant charge.}
\end{figure}

The second test consisted of a more astronomically realistic exposure profile : 6 up-the-ramp exposures each of 600s and followed by a reset. The illumination source was a very faint LED with constant flux over the course of the exposure.  The result is shown in Fig.~\ref{fig:compareplot3}. Finally, 40 up-the-ramp exposures, shown in Fig.~\ref{fig:compareplot2}, corresponding to a half-night of observation  were used. In both cases there was a good fit between the model and the data.

\begin{figure}[htbp]
\begin{center}
\begin{tabular}{c}
\includegraphics[height=9cm]{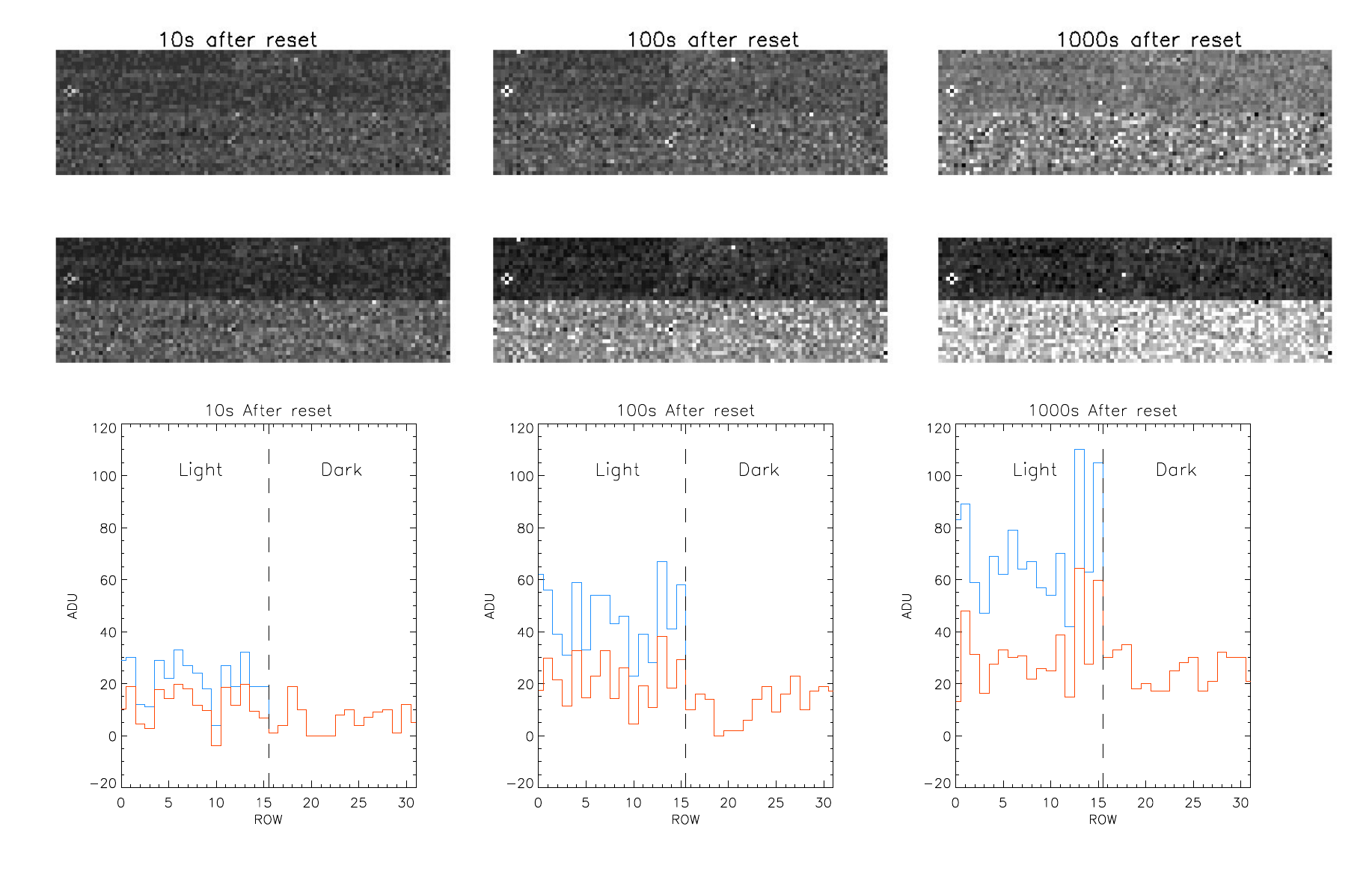}
\end{tabular}
\end{center}
\caption
{ \label{fig:LongSoakTestallinfo}
Persistence corrected images in top row, raw images in bottom row. The initial exposure (applied only to the
lower half of the image) consisted of a single 16384s soak at approximately 30ke\(^-\) with the ERIS 5.3\(\mu\)m detector.
Vertical cuts through raw images shown in blue, through corrected images in orange. The gain was 4.4e\(^-\)/ADU.}
\end{figure}

\begin{figure}[htbp]
\begin{center}
\begin{tabular}{c}
\includegraphics[height=9cm]{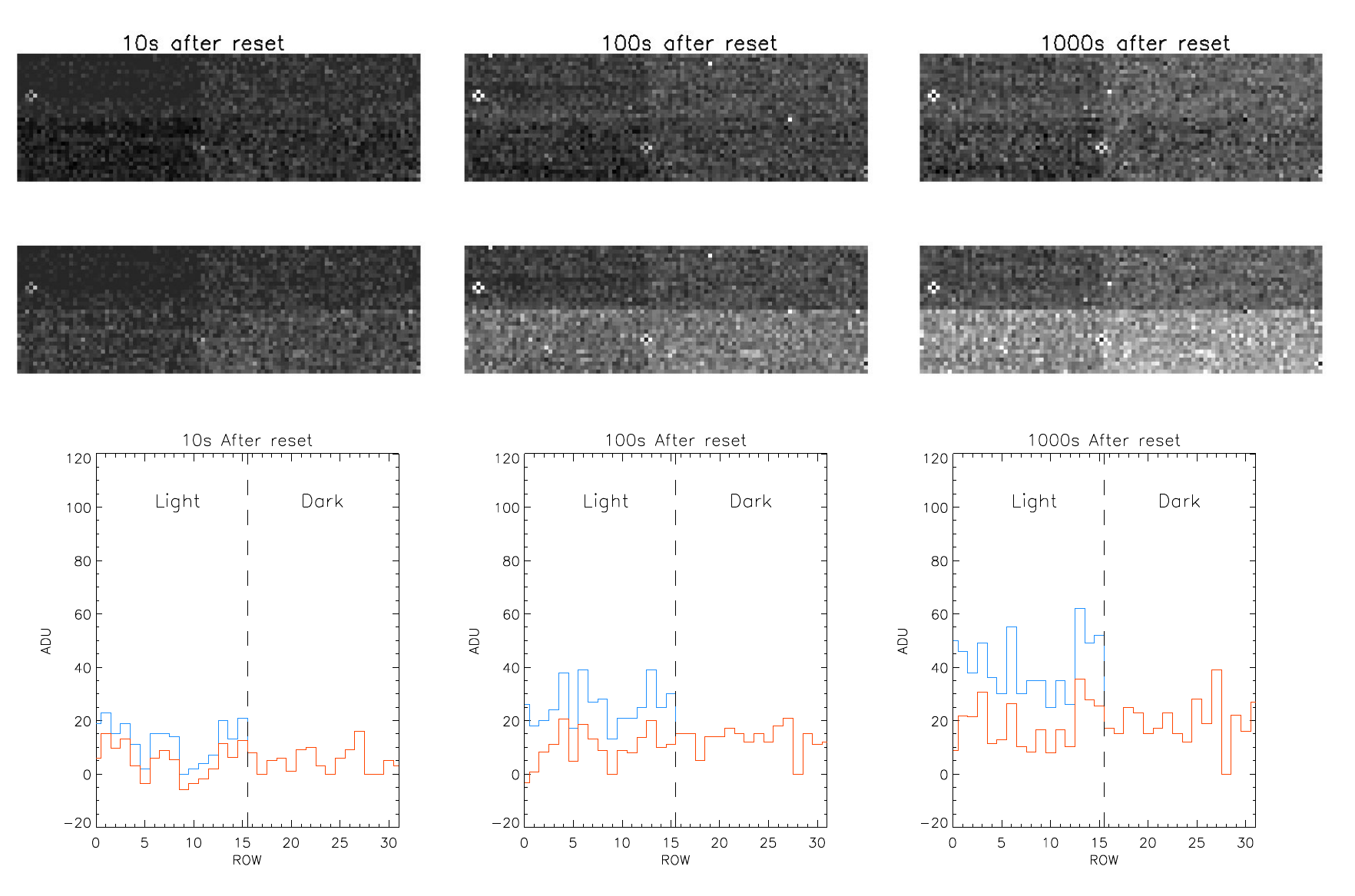}
\end{tabular}
\end{center}
\caption
{ \label{fig:RampTest3allinfo}
As for Fig.~\ref{fig:LongSoakTestallinfo}. except that the measurements directly follow a series of twelve 600s up-the-ramp exposures to approximately 30ke\(^-\).}
\end{figure}

The tests up to this point were done on small windows.  This improved the SNR since the response of many pixels was averaged together. The next test was done on a pixel by pixel basis with the results presented as corrected image sequences. This check ensured that the persistence removal processing did not introduce noise artifacts. A small region of the detector measuring 100 x 32 pixels was chosen in an area where the stray light was at a minimum. This region was weakly illuminated by an LED such that a fluence of 30ke\(^-\) was reached in 600s. Whenever the LED was illuminated the top part of the window was rapidly reset line-by-line in a 20ms loop. This defined a persistence-free dark reference area which served to highlight the efficiency of the model. In the first test the LED was on for a single 600s period after which the pixels were left to soak for 16384s. The detector was then reset and the detrapped charge measured using a series of non-destructive reads for a further 1000s. These exposures were assembled into a movie and compared side by side with a second movie where the frames had the output of the persistence model subtracted from them. Stills from the movie are shown in Figs.~\ref{fig:LongSoakTestallinfo}. Below each pair of stills is a plot showing a vertical cut through each image. It seems that the noise in the corrected image did not increase significantly. As well as being dark-subtracted, the windowed pixels were also reference pixel subtracted in both x and y axes. A second similar test was then performed but replacing the single long soak by a series of twelve up-the-ramp 600s exposures once again followed by a reset and a series of non-destructive reads. The corresponding movie stills are shown in Fig.~\ref{fig:RampTest3allinfo}. Once again the images were well corrected.

\section{Conclusions}
\label{sect:concs}  
We have developed a method to characterize the behaviour and number of traps in HxRG detectors, and developed a model that describes how the charging and discharging of these traps can result in reciprocity failure and persistence, respectively. Our depletion region model has extended the previous model presented by Smith\cite{SMITH1} to include a hypothesis of how the charge-trapping process can also be explained as movement of electrons across the depletion edge. The effective fractional charge of these moving electrons has also been measured.
The good agreement between the predictive model and the measured persistence data from three H2RG detectors indicates that our  model describes the behaviour of the trapped charges at a level adequate for useful correction. Our mathematical model is an advantage over previous ones because it calculates the detrapping charge profile as a function of both the previous exposure level and also the previous exposure time using IIR filters. This model can seamlessly handle any arbitrary historical exposure profile.

We encountered a few surprises in our study. The result that the traps take very much longer to charge than they do to discharge was unexpected. Future work could attempt to explain this energetically. For example, to fill a trap requires an electron to acquire sufficient thermal energy to move against the potential to reach DR trap sites. Subsequent detrapping is more energetically favorable since the trapped charge only needs to acquire sufficient energy to move out of the local potential well of the trap (which could be relatively shallow) before being swept back to the \textit{n} region by the intrinsic electric field of the DR. Lower operating temperatures would reduce the likelihood of electrons reaching the DR traps and so reduce the amount of trapped charge. Cooling has already been observed\cite{TULLOCH} to reduce persistence.

\section{Future Work}
\label{sect:future}  
There are several other interesting avenues to explore that could further increase the accuracy of the model. Further work could also attempt to include the behaviour of trapped holes, which has been entirely ignored in the current study. The effect of gaps in the exposure record on the persistence engine accuracy also needs to be investigated. In an observatory environment it will require great care to ensure that all exposures produce an image file that can be used to calculate the trapped charge. Inadvertent exposures or ones that reach full-well within a small fraction of the exposure time will cause errors. Further work needs to be done to determine whether the persistence in neighboring pixels are affected by saturated pixels, as was seen by Crouzet et al.\cite{Crouzet}. The persistence maps presented here were obtained using 60s soaks followed by 60s darks to collect the detrapped charge. This meant that the maps contained primarily short-time constant trapped charge. It would be interesting to repeat these measurements using longer soaks and dark exposures.

The next step is to implement the persistence engine in an astronomical instrument. Two approaches to this implementation are possible. In the ESO observatories, data acquisition computers (known as LLCUs) already process detector data streams using Fowler and up-the-ramp H2RG sequences as they arrive from the detector controller. The LLCUs have sufficient capacity to also implement the persistence engine code. Using the internal trapped-charge images in each of the five bins, persistence frames could be generated for each exposure in near real-time and appended as image extensions to the scientific data. It would then be up to the astronomer to make use of the persistence frames or not. Alternatively, a persistence map could be created post-facto using data stored in the archive, however this requires that even all engineering images are saved to the ESO archive. In any case saving every exposure is recommended, as it would allow the persistence maps to be regenerated in the event of future updates to the model.

ESO plans to test this persistence correction method with future instruments, and the on-sky performance can then be evaluated.

The authors can be contacted for copies of the IDL code used in this study.



\acknowledgments
Thanks to Leander Mehrgan for providing the NGC detector controller electronics and Christopher Mandla for many useful suggestions on the persistence measurement process.


\bibliography{article}   
\bibliographystyle{spiejour}   


\vspace{2ex}\noindent\textbf{Simon Tulloch} is a detector engineer at the European Southern Observatory. He received his BS degree in Physics from the University of Kent and his PhD degree in Astrophysics from the University of Sheffield in 2010. He was previously head of the detector group at the Isaac Newton Group of telescopes in La Palma, Spain. His research interests include low-noise signal processing and photon counting detectors.

\vspace{1ex}\noindent\textbf{Elizabeth M. George} is a detector engineer in the Detector Systems Group at European Southern Observatory. She develops instrumentation and detectors for astronomical telescopes from the visible to millimeter wavelengths. Previously, she completed her postdoctoral work on the ERIS project at the Max Planck Institute for Extraterrestrial Physics and her doctoral work in physics at UC Berkeley on the South Pole Telescope experiment.

\listoffigures
\listoftables

\end{spacing}
\end{document}